\documentclass[12pt]{article}
\pdfoutput=1

\usepackage{amsmath}
\usepackage{amssymb}
\usepackage{epsfig}
\usepackage{epstopdf}
\usepackage{latexsym}
\numberwithin{equation}{section}
\usepackage[
      colorlinks=false,
      linkcolor=darkblue,  
      urlcolor=blue,    
      filecolor=blue,     
      citecolor=red,
linktocpage=true,
      pdfstartview=FitV,
      bookmarksopen=true    
      ]{hyperref}
\begin{document}

\begin{titlepage}

\centerline{\Huge \rm Supersymmetric Janus solutions} 
\bigskip
\centerline{\Huge \rm  in five and ten dimensions}
\bigskip
\bigskip
\bigskip
\bigskip
\bigskip
\bigskip
\centerline{\rm Minwoo Suh}
\bigskip
\centerline{\it Department of Physics and Astronomy} 
\centerline{\it University of Southern California} 
\centerline{\it Los Angeles, CA 90089, USA}
\bigskip
\centerline{\tt minsuh@usc.edu} 
\bigskip
\bigskip
\bigskip
\bigskip
\bigskip

\begin{abstract}
We explicitly truncate $\mathcal{N}$ = 8 gauged supergravity in five dimensions to its $SU(3)$-invariant sector with dilaton and axion fields. We show that this truncation has a solution which is identical to the super Janus constructed in $\mathcal{N}$ = 2 gauged supergravity in five dimensions. Then we lift the solution of the $SU(3)$-invariant truncation to type IIB supergravity by employing the consistent truncation ansatz. We show that the lifted solution falls into a special case of the supersymmetric Janus solutions constructed in type IIB supergravity. Additionally, we also prove that the lifted solution provides a particular example of the consistent truncations of type IIB supergravity on Sasaki-Einstein manifolds. 
\end{abstract}

\bigskip
\bigskip
\bigskip
\vskip 7cm

\flushleft {July, 2011}

\end{titlepage}

\tableofcontents

\vskip 4cm

\section {Introduction}

The Janus solutions provide a class of examples for the AdS/CFT correspondence \cite{Maldacena:1997re}. The Janus solutions are characterized by two main features: (i) they are $AdS$-domain wall solutions with an interface, (ii) the dilaton field takes constant values on both sides of the interface, but it jumps across the interface. As the dilaton field is not constant, the coupling constant of the dual gauge theory varies across the interface, {\it i.e.} the dual gauge theories are defect conformal field theories. The first example of Janus solutions was discovered in type IIB supergravity with no supersymmetries by Bak, Gutperle and Hirano in \cite{Bak:2003jk}. The dual gauge theory is $\mathcal{N}$ = 4 super Yang-Mills theory in 3+1 dimensions with a 2+1 dimensional interface. Even though this solution breaks all the supersymmetries, the stability against a large class of perturbations was proved in \cite{Bak:2003jk, Freedman:2003ax}.

After the discovery of the original Janus solution, the dual gauge theory was studied in \cite{Clark:2004sb}. It was observed that by reducing $SO(6)$ R-symmetry of the dual gauge theory down to at least $SU(3)$, some supersymmetries were restored. Motivated by this observation, Clark and Karch constructed a supersymmetric Janus solution with $SU(3)$ isometry, super Janus \cite{Clark:2005te}, based on the studies of curved domain wall solutions \cite{LopesCardoso:2001rt, Chamseddine:2001ga, Chamseddine:2001hx, LopesCardoso:2002ec, Behrndt:2002ee} in $\mathcal{N}$ = 2 gauged supergravity with one hypermultiplet in five dimensions \cite{Ceresole:2000jd, Ceresole:2001wi}.

Later, Janus gauge theories were constructed more systematically in \cite{D'Hoker:2006uv}. It gives the complete classification of all possible Janus solutions in type IIB supergravity. According to the classification, there are four kinds of solutions with $SO(6)$, $SU(3)$, $SU(2){\times}U(1)$ and $SO(3){\times}SO(3)$ isometries, and each of them has zero, four, eight, and sixteen Poincar\'e supersymmetries, respectively. Among these, the Janus solution with no supersymmetry is the original Janus solution \cite{Bak:2003jk}. By D'Hoker, Estes and Gutperle, the Janus solutions with four and sixteen supersymmetries were constructed in type IIB supergravity in \cite{D'Hoker:2006uu} and \cite{D'Hoker:2007xy, D'Hoker:2007xz}, respectively. Later, the Janus field theories in \cite{D'Hoker:2006uv} were generalized to allow the theta-angle to vary which is holographicallly dual to the axion field, and were also applied to construct three-dimensional Chern-Simons theories with $\mathcal{N}$ = 4 supersymmetries in \cite{Gaiotto:2008sd}. 

Despite of all these developments in Janus geometries, as the five- and ten-dimensional solutions were constructed independently, the relation between those solutions are far from obvious. However, as $\mathcal{N}$ = 2 gauged supergravity with one hypermultiplet is a truncation of $\mathcal{N}$ = 8 gauged supergravity in five dimensions \cite{Pernici:1985ju, Gunaydin:1984qu, Gunaydin:1985cu}, it was conjectured by Clark and Karch in \cite{Clark:2004sb} that the super Janus in $\mathcal{N}$ = 2 gauged supergravity would be embedded in $\mathcal{N}$ = 8 gauged supergravity in five dimensions. If this embedding could be achieved, as the lift of $\mathcal{N}$ = 8 gauged supergravity to type IIB supergravity is readily known \cite{Khavaev:1998fb, Pilch:2000ue, Pilch:2000fu}, one should be able to uplift the supersymmetric Janus solution in five dimensions to the one in IIB. This will provide us with the bridge between the known supersymmetric Janus solutions in five and ten dimensions.

In order to address these questions, we will revisit the $SU(3)$-invariant truncation of $\mathcal{N}$ = 8 gauged supergravity in five dimensions which was studied in \cite{Gunaydin:1985cu} and \cite{Girardello:1998pd, Distler:1998gb, Girardello:1999bd}. Later it was uplifted to type IIB supergravity in \cite{Pilch:2000fu}. However, in these studies, there was only one real scalar field in the flat domain wall, and the dilaton/axion fields were suppressed. In order to construct Janus solutions, we will generalize the previous studies in two aspects: (i) we extend the field content to include the dilaton/axion fields, so we will have two complex or four real scalar fields, (ii) we consider the $AdS$-domain wall instead of the flat domain wall. However, as it was known in $\mathcal{N}$ = 2 gauged supergravity in five dimensions in \cite{LopesCardoso:2001rt, LopesCardoso:2002ec}, we will find that the two directions of generalization are in fact equivalent, {\it i.e. one can turn on the dilaton/axion fields only in the curved background, and vice versa}. Finally we will show that the $SU(3)$-invariant truncation with the dilaton/axion fields indeed has a solution identical to the super Janus in \cite{Clark:2005te}.

Then we will uplift the solution of the $SU(3)$-invariant truncation to type IIB supergravity by employing the consistent truncation ansatz for metric and dilaton/axion fields in \cite{Khavaev:1998fb, Pilch:2000ue, Pilch:2000fu}. Though there are the lift formulae for three- and five-form fluxes proposed in \cite{Khavaev:2001yg}, we find that they do not work for the curved domain walls. We propose modified lift formulae similar to those of  \cite{Khavaev:2001yg} for three- and five-form fluxes, and check that they generate correct fluxes for the cases we are considering. Finally we will show that the lift of the $SU(3)$-invariant truncation indeed falls into a special case of the supersymmetric Janus solution in type IIB supergravity in \cite{D'Hoker:2006uu}. 

Of independent interest from the Janus solutions, there has been notable development in consistent truncation of type IIB supergravity on Sasaki-Einstein manifolds recently \cite{Cassani:2010uw, Gauntlett:2010vu, Liu:2010sa, Skenderis:2010vz}. We will show that the lift of the $SU(3)$-invariant truncation to type IIB supergravity provides a particular example of the truncation in \cite{Cassani:2010uw, Gauntlett:2010vu}. 

In section 2 we begin by studying the $SU(3)$-invariant truncation of $\mathcal{N}$ = 8 gauged supergravity in five dimensions with dilaton and axion fields. In section 3 we show that a solution of the $SU(3)$-invariant truncation is identical to the super Janus in $\mathcal{N}$ = 2 supergravity in five dimensions. In section 4 we lift the solution of the $SU(3)$-invariant truncation to type IIB supergravity by employing consistent truncation ansatz for metric and dilaton/axion fields. In section 5 we show that the lifted metric and dilaton/axion fields completely fixes the supersymmetric Janus solution with $SU(3)$ isometry in type IIB supergravity. In section 6 we continue the lift of the $SU(3)$-invariant truncation for three- and five-form fluxes. In section 7 we consider the consistent truncation of type IIB supergravity on Sasaki-Einstein manifolds in relation with the $SU(3)$-invariant truncation. In section 8 concluding remarks are offered. In appendix A we briefly review $\mathcal{N}$ = 8 gauged supergravity in five dimensions. In appendix B the $SU(2,1)$ algebra is presented. In appendix C details of the supersymmetry variation for spin-3/2 fields are presented for the $SU(3)$-invariant truncation. Appendix D summarizes the different parametrizations of the scalar manifold in this paper. In appendix E we present the field equations in five dimensions.

\section{Truncation of $\mathcal{N}$ = 8 gauged supergravity in five dimensions}

\subsection{The $SU(3)$-invariant truncation}

We study the $SU(3)$-invariant truncation of $\mathcal{N}$ = 8 gauged supergravity in five dimensions. There are a graviton $e_{\mu}\,^a$, a vector field $A_{\mu}$, and four real scalars $x_{i}$ for the bosonic field content in the $SU(3)$-invariant sector. As mentioned in the introduction, there have been studies on the $SU(3)$-invariant truncation in \cite{Gunaydin:1985cu, Girardello:1998pd, Distler:1998gb, Girardello:1999bd} and \cite{Pilch:2000fu}, but there was only one real scalar field as a consistent subsector in these studies. Here we extend the field content to all four scalar fields including dilaton and axion fields.

The 42 scalar fields of $\mathcal{N}$ = 8 gauged supergravity in five dimensions live on the coset manifold $E_{6(6)}/USp(8)$. The basic structure of the coset manifold is explained in \cite{Gunaydin:1985cu}, and is summarized in appendix A. Fundamental representation of $E_{6(6)}$ is real and 27-dimensional. The infinitesimal $E_{6(6)}$ transformation in the $SL(6,\mathbb{R})$${\times}$$SL(2,\mathbb{R})$ basis, ($z_{IJ}$, $z^{I{\alpha}}$), is \cite{Gunaydin:1985cu}
\begin{align} \label{zls1}
{\delta}z_{IJ}\,=\,&-\Lambda^K\,_I\,z_{KJ}\,-\Lambda^K\,_J\,z_{IK}\,+\Sigma_{IJK\beta}\,z^{K\beta}\,, \\ \label{zls2}
{\delta}z_{I{\alpha}}\,=\,&\,\,\Lambda^I\,_K\,z^{K\alpha}\,+\,\Lambda^{\alpha}\,_{\beta}\,z_{I{\beta}}\,+\,\Sigma^{KLI\beta}\,z_{KL}\,,
\end{align}
where $\Lambda^I\,_J$ and $\Lambda^\alpha\,_\beta$ are real and traceless generators of $SL(6,\mathbb{R})$ and $SL(2,\mathbb{R})$ respectively, and $\Sigma_{IJK{\alpha}}$ is real and antisymmetric in $IJK$.

Among the $E_{6(6)}$ generators, the $SU(3)$ generators of the gauge group $SO(6)$ are the ones that commute with the complex structure, $J_{IJ}$, which is an antisymmetric tensor with nonzero components, $J_{12}\,=\,J_{34}\,=\,J_{56}\,=\,1$. Then we obtain the $SU(3)$-invariant generators by finding ones which commute with the $SU(3)$ generators. There are eight $SU(3)$-invariant generators, and they close onto an $SU(2,1)$ algebra,
\begin{align}
\Sigma_{IJK{\alpha}}^{(1)}&=+(\delta_{IJK{\alpha}}^{1\,3\,5\,7}-\delta_{IJK{\alpha}}^{2\,4\,6\,8})+(\delta_{IJK{\alpha}}^{1\,3\,6\,8}-\delta_{IJK{\alpha}}^{2\,4\,5\,7})+(\delta_{IJK{\alpha}}^{1\,4\,5\,8}-\delta_{IJK{\alpha}}^{2\,3\,6\,7})-(\delta_{IJK{\alpha}}^{1\,4\,6\,7}-\delta_{IJK{\alpha}}^{2\,3\,5\,8}), \ \\
\Sigma_{IJK{\alpha}}^{(2)}&=+(-\delta_{IJK{\alpha}}^{1\,3\,5\,8}-\delta_{IJK{\alpha}}^{2\,4\,6\,7})+(\delta_{IJK{\alpha}}^{1\,3\,6\,7}+\delta_{IJK{\alpha}}^{2\,4\,5\,8})+(\delta_{IJK{\alpha}}^{1\,4\,5\,7}+\delta_{IJK{\alpha}}^{2\,3\,6\,8})-(-\delta_{IJK{\alpha}}^{1\,4\,6\,8}-\delta_{IJK{\alpha}}^{2\,3\,5\,7}), \\
\Sigma_{IJK{\alpha}}^{(3)}&=+(\delta_{IJK{\alpha}}^{1\,3\,5\,8}-\delta_{IJK{\alpha}}^{2\,4\,6\,7})+(\delta_{IJK{\alpha}}^{1\,3\,6\,7}-\delta_{IJK{\alpha}}^{2\,4\,5\,8})+(\delta_{IJK{\alpha}}^{1\,4\,5\,7}-\delta_{IJK{\alpha}}^{2\,3\,6\,8})-(\delta_{IJK{\alpha}}^{1\,4\,6\,8}-\delta_{IJK{\alpha}}^{2\,3\,5\,7}), \\
\Sigma_{IJK{\alpha}}^{(4)}&=+(\delta_{IJK{\alpha}}^{1\,3\,5\,7}+\delta_{IJK{\alpha}}^{2\,4\,6\,8})+(-\delta_{IJK{\alpha}}^{1\,3\,6\,8}-\delta_{IJK{\alpha}}^{2\,4\,5\,7})+(-\delta_{IJK{\alpha}}^{1\,4\,5\,8}-\delta_{IJK{\alpha}}^{2\,3\,6\,7})-(\delta_{IJK{\alpha}}^{1\,4\,6\,7}+\delta_{IJK{\alpha}}^{2\,3\,5\,8}), 
\end{align}
\begin{align}
\Lambda^{(5)\,I}\,_J\,&=\,J_{IJ}, \\
\Lambda^{(6)\,\alpha}\,_\beta\,&=\,(S_1)\,^\alpha\,_\beta, \\
\Lambda^{(7)\,\alpha}\,_\beta\,&=\,(S_2)\,^\alpha\,_\beta, \\
\Lambda^{(8)\,\alpha}\,_\beta\,&=\,(S_3)\,^\alpha\,_\beta,
\end{align}
where
\begin{equation}
S_1\,=\,\left(
\begin{array}{ll}
 0 & -1 \\
 -1 & 0
\end{array}
\right)\,,\,\,\,\,
S_2\,=\,\left(
\begin{array}{ll}
 1 & 0 \\
 0 & -1
\end{array}
\right)\,,\,\,\,\,
S_3\,=\,\left(
\begin{array}{ll}
 0 & 1 \\
 -1 & 0
\end{array}
\right)\,,
\end{equation}
are three $SL(2,\mathbb{R})$ generators. We refer to appendix B for the $SU(2,1)$ algebra of these generators. The generators $\Lambda^{(6)}$, $\Lambda^{(7)}$ are symmetric, and with the self-duality defined by
\begin{equation}
\Sigma_{IJK{\alpha}}\,=\,+\,\frac{1}{6}\,\epsilon_{IJKLMNP}\,\epsilon_{\alpha\beta}\,\Sigma^{MNP\beta}\,,
\end{equation}
$\Sigma^{(1)}$, $\Sigma^{(2)}$ are self-dual. 
By computing the Cartan-Killing form \cite{Gunaydin:1985cu} these symmetric and self-dual generators turn out to be the noncompact generators of the scalar manifold \cite{Ceresole:2001wi},
\begin{equation}
\mathcal{M}\,=\,\frac{SU(2,1)}{SU(2){\times}U(1)}\,.
\end{equation}

We exponentiate the transformations by four noncompact generators, 
\begin{align}
T_1\,&=\,{\frac{1}{4\sqrt{2}}}\,\Sigma^{(1)}\,, \,\,\,\,\,\,\,\,\,\,\,\,\, \,\,\,\,\,\,\,\,\,\, \,\,\,\,\,\,\,\,\,\,  T_2\,=\,{\frac{1}{4\sqrt{2}}}\,\Sigma^{(2)}\,, \,\,\,\,\,\,\,\,\,\, \notag \\ T_3\,&=\,{\frac{1}{2\sqrt{2}}}\,(\Lambda^{(7)}\,+\,\Lambda^{(6)})\,, \,\,\,\,\,\,\,\,\,\, T_4\,=\,{\frac{1}{2\sqrt{2}}}\,(\Lambda^{(7)}\,-\,\Lambda^{(6)})\,,
\end{align}
with parameters, $x_1$, $x_2$, $x_3$, $x_4$, respectively. Schematically the exponentiation of the generators is
\begin{equation} \label{exp}
z'\,=\,e^{(x_3\,T_3\,+\,x_4\,T_4)}\,e^{(x_1\,T_1\,+\,x_2\,T_2)}\,z\,.
\end{equation}
From the exponentiation we can extract the coset representatives in the $SL(6,\mathbb{R})$${\times}$$SL(2,\mathbb{R})$ basis, $U^{IJ}\,_{KL}$, $U^{IJK{\alpha}}$, $U_{I\alpha}\,^{KL}$ and $U_{I{\alpha}}\,^{J{\beta}}$, by \eqref{zU1} and \eqref{zU2}. The coset representatives in the $USp(8)$ basis, $\mathcal{V}^{IJab}$, $\mathcal{V}_{I{\alpha}}\,^{ab}$, are obtained by \eqref{UV1} and \eqref{UV2}.

Now with the coset representatives in the $USp(8)$ basis, we can reduce the Lagrangian of the $SU(3)$-invariant truncation. We introduce an angular parametrization of the scalar fields,
\begin{align} \label{su3para}
x_1\,&=\,2\,\chi\,\cos\psi\,, \,\,\,\, x_2\,=\,2\,\chi\,\sin\psi\,, \notag \\ x_3\,&=\,2\,\phi\,\cos{a}\,, \,\,\,\,\, x_4\,=\,2\,\phi\,\sin{a}\,.
\end{align}
The bosonic part of the Lagrangian is
\begin{equation} \label{su3lag}
e^{-1}\,\mathcal{L}\,=\,-\,\frac{1}{4}\,R\,+\,\mathcal{L}_{kin}\,+\,\mathcal{P}\,-\,\frac{3}{4}\,F_{\mu\nu}\,F^{\mu\nu}\,+\,\mathcal{L}_{CS}\,,
\end{equation}
where the kinetic term for the scalar fields is
\begin{align} \label{su3kin}
\mathcal{L}_{kin}\,\,=\,\,&\frac{1}{2}\,\partial_\mu\chi\,\partial^\mu\chi\,+\,\frac{1}{8}\,\sinh^2(2\,\chi)\,\left(\partial_\mu\psi\,+\,\sinh^2\phi\,\partial_\mu{a}\,+\,g\,A_\mu\right)^2 \notag \\ +\,&\cosh^2\chi\,\left(\frac{1}{2}\,\partial_\mu\phi\,\partial^\mu\phi\,+\,\frac{1}{8}\,\sinh^2{(2\,\phi)}\,\partial_\mu{a}\,\partial^\mu{a}\right)\,,
\end{align}
and the scalar potential is
\begin{equation}
\mathcal{P}\,=\,\frac{3}{32}\,g^2\,\Big(\cosh^2(2\,\chi)\,-\,4\,\cosh(2\,\chi)\,-\,5\,\Big)\,.
\end{equation}
{\it Note that the scalar potential is manifestly invariant under $SL(2,\mathbb{R})$, ${\it i.e.}$ it is independent of $\phi$ and $a$. We note that $\phi$ and $a$ are dilaton and axion fields in five dimensions.}

The scalar potential admits two critical points which are the $AdS_5$ vacua in the $SU(3)$-invariant truncation \cite{Khavaev:1998fb, Freedman:1999gp}.{\footnote { The scalar field $\chi$  was denoted by $\varphi_1$ = $\chi$ in \cite{Freedman:1999gp}.}} One of the critical points is the $\mathcal{N}$ = 8 supersymmetric $SO(6)$ point where $\chi$ = 0 and $\mathcal{P}$ = $-\,\frac{3}{4}\,g^2$. This point lifts to $AdS_5\,\times\,S^5$ vacua in type IIB supergravity.  Another one is the nonsupersymmtric $SU(3)$ point where $\chi$ = $\frac{1}{2}\,\log(2-\sqrt{3})$  and $\mathcal{P}$ = $-\,\frac{27}{32}\,g^2$. This point lifts to a solution found by Romans in type IIB supergravity in \cite{Romans:1984an}. The holographic renormalization flows studied in \cite{Girardello:1998pd, Distler:1998gb, Girardello:1999bd, Pilch:2000ue} and the domain wall solution for holographic superconductor in \cite{Gubser:2009qm, Gubser:2009gp} flow to this critical point. 

Before we finish this section, let us count the number of bosonic fields in the $SU(3)$-invariant truncation. In the full theory, under the gauge group, $SU(4)\,{\simeq}\,SO(6)$, 1 graviton $e_{\mu}\,^a$ transforms as $\bf1$, 15 vector fields $A_{{\mu}IJ}$ as $\bf15$, 12 two-form tensor fields  $B_{\mu\nu}\,^{I{\alpha}}$ as $\bf6+6$, and 42 scalar fields $\phi^{abcd}$ as $\bf20'+10+\overline{10}+1+1$. By breaking $SU(4)$ down to $SU(3)$ they branch as \cite{Distler:1998gb}
\begin{align}
e_{\mu}\,^a \,\,\,\,\,\,\,\,\,\,\,\,\,\,\,\,\,\,\,\,\,\,\,\,\,\,\,\,\,\,\,\,\,\,\,\,\,\,\,\,\,\, \bf1\,\,\,\,\,\,\,\,\,\,\,\,\,\,\,\,\,\,\,\,&{\rightarrow}\,\,\,\,\,\,\,\,\,\,\,\,\,\,\,\,\,\,\,\,\bf1, \\
A_{\mu{IJ}} \,\,\,\,\,\,\,\,\,\,\,\,\,\,\,\,\,\,\,\,\,\,\,\,\,\,\,\,\,\,\,\,\,\,\,\, \bf15\,\,\,\,\,\,\,\,\,\,\,\,\,\,\,\,\,\,\,\,&{\rightarrow}\,\,\,\,\,\,\,\,\,\,\,\,\,\,\,\,\,\,\,\,\bf8+3+\overline3+1, \\
B_{\mu\nu}^{I\alpha} \,\,\,\,\,\,\,\,\,\,\,\,\,\,\,\,\,\,\,\,\,\,\,\,\,\,\,\,\,\, \bf6+6\,\,\,\,\,\,\,\,\,\,\,\,\,\,\,\,\,\,\,\,&{\rightarrow}\,\,\,\,\,\,\,\,\,\,\,\,\,\,\,\,\,\,\,\,\bf(3+\overline3)+(3+\overline3), \\
\bf20'\,\,\,\,\,\,\,\,\,\,\,\,\,\,\,\,\,\,\,\,&{\rightarrow}\,\,\,\,\,\,\,\,\,\,\,\,\,\,\,\,\,\,\,\,\bf8+6+\overline6, \notag\\
\phi^{abcd} \,\,\,\,\,\,\,\,\,\,\,\,\,\,\,\,\,\,\,\, \bf10+\overline{10}\,\,\,\,\,\,\,\,\,\,\,\,\,\,\,\,\,\,\,\,&{\rightarrow}\,\,\,\,\,\,\,\,\,\,\,\,\,\,\,\,\,\,\,\,\bf(1+\overline3+6)+(1+3+\overline6), \\
\bf1+1\,\,\,\,\,\,\,\,\,\,\,\,\,\,\,\,\,\,\,\,&{\rightarrow}\,\,\,\,\,\,\,\,\,\,\,\,\,\,\,\,\,\,\,\,\bf1+1, \notag
\end{align}
so we have a graviton $e_{\mu}\,^a$, a vector field $A_{\mu}$, and four scalars $x_{i}$ in the $SU(3)$-invariant sector.

\subsection{The supersymmetry equations}

In this subsection we will explicitly derive the supersymmetry equations for the $SU(3)$-invariant truncation with the dilaton and axion fields, and then solve them numerically. Some equivalent equations in $\mathcal{N}$ = 2 gauged supergravity were obtained in \cite{LopesCardoso:2001rt, LopesCardoso:2002ec}, however, this subsection is to have equations in the parametrization of $\mathcal{N}$ = 8 gauged supergravity in five dimensions.

We will consider the $AdS$-domain wall,
\begin{equation} \label{AdSmet}
ds^2\,=\,e^{2\,U(r)}\,ds^2_{AdS_4}\,+\,dr^2\,,
\end{equation}
where
\begin{equation} \label{ads4}
ds^2_{AdS_4}\,=\,e^{2z/l}\,(+\,dt^2\,-\,dx^2\,-\,dy^2)\,-dz^2\,.
\end{equation}
When $l\,\rightarrow\,\infty$, it reduces to the flat domain wall.

We begin by considering the superpotential and the spinors in five dimensions. The superpotential, $W$, is obtained as one of the eigenvalues of $W_{ab}$ tensor \cite{Freedman:1999gp},
\begin{equation}
W_{ab}\,\eta^b_{(k)}\,=\,W\,\eta^a_{(k)}\,,
\end{equation}
where $k\,=\,1,\,2$. There are two eigenvalues with degeneracy of two and six, and they are, respectively,
\begin{align} \label{sp}
W_1\,&=\,-\,\frac{3}{4}\,\Big(1\,+\,\cosh(2\,\chi)\Big)\,, \\
W_2\,&=\,-\,\frac{1}{4}\,\Big(5\,+\,\cosh(2\,\chi)\Big)\,,
\end{align}
but only $W\,=\,W_1$ gives the scalar potential by
\begin{equation}
\mathcal{P}\,=\,\frac{g^2}{8}\,\left|\frac{\partial{W}}{\partial\varphi_i}\right|^2\,-\,\frac{g^2}{3}|W|^2\,,
\end{equation}
where $\varphi_i\,=\,\chi\,,\,\phi\,,\,\psi\,,\,a\,.$
The eigenvectors, $\eta^a_{(1)}$, $\eta^a_{(2)}$, for the superpotential, $W$, are
\begin{align}
\eta^a_{(1)}\,&=\,(0,\,1,\,0,\,1,\,-1,\,0,\,1,\,0)\,, \\
\eta^a_{(2)}\,&=\,(-1,\,0,\,1,\,0,\,0,\,-1,\,0,\,-1)\,,
\end{align}
and they are related to each other by
\begin{equation}
\Omega_{ab}\,\eta^b_{(1)}\,=\,-\,\eta^a_{(2)},\,\,\,\,\,\,\,\,\,\, \Omega_{ab}\,\eta^b_{(2)}\,=\,+\,\eta^a_{(1)}\,,
\end{equation}
where $\Omega_{ab}$ is the $USp(8)$ symplectic form given in {\it e.g.} \cite{Freedman:1999gp}. We employed the gamma matrix conventions in \cite{Freedman:1999gp}. Then the $SU(3)$-invariant five-dimensional spinors are defined by
\begin{align}
\epsilon^a\,&=\,\eta^a_{(1)}\,\hat{\epsilon}_1\,+\,\eta^a_{(2)}\,\hat{\epsilon}_2\,, \\
\epsilon_a\,&=\,\Omega_{ab}\,\epsilon^b\,=\,-\,\eta^a_{(2)}\,\hat{\epsilon}_1\,+\,\eta^a_{(1)}\,\hat{\epsilon}_2\,,
\end{align}
where $\hat{\epsilon}_1$ and $\hat{\epsilon}_2$ are spinors with four complex components.

The supersymmetry equations are obtained by setting the supersymmetry variations of fermionic fields, {\it i.e.} the spin-3/2 and spin-1/2 fields, to zero. For the supersymmetry analysis we will suppress the gauge field, $A_\mu$, below. The bosonic parts of the variations are \cite{Gunaydin:1985cu}
\begin{align} \label{gravvar}
\delta\,\psi_{\mu{a}}\,&=\,D_\mu\,\epsilon_a\,-\,\frac{1}{6}\,g\,W_{ab}\,\gamma_\mu\,\epsilon^b\,, \\ \label{s2}
\delta\,\chi_{abc}\,&=\,\sqrt{2}\,\Big[\gamma^\mu\,P_{\mu{abcd}}\,\epsilon^d\,-\,\frac{1}{2}\,g\,A_{dabc}\,\epsilon^d\Big]\,.
\end{align}
First we solve the spin-3/2 field variation. For the $t$-, $x$-, $y$- directions, \eqref{gravvar} reduces to
\begin{equation} \label{vartxy}
U'\,\gamma^{(4)}\,\epsilon_a\,+\,e^{-U}\,\gamma^{(3)}\,\epsilon_a\,-\frac{1}{3}\,g\,W_{ab}\,\epsilon^b\,=\,0\,.
\end{equation}
From the integrability of the variation, we obtain \cite{LopesCardoso:2001rt, LopesCardoso:2002ec}
\begin{equation} \label{su3U}
U'\,=\,\mp\,\frac{1}{3}\,g\,W\,\gamma,
\end{equation}
where
\begin{equation}
\gamma\,=\,\sqrt{1\,-\frac{9\,e^{-2U}}{l^2\,g^2\,W^2}}\,,
\end{equation}
and the prime is a derivative with respect to the $r$-coordinate. From here the upper and lower signs in equations are related. Note that for the flat domain wall, $l\,\rightarrow\,\infty$, we have $\gamma\,=\,1$. We also obtain a projection condition for the spinors,
\begin{equation} \label{proj1}
\hat{\epsilon}_1\,=\,+\,(\mp\,\gamma\,\gamma^{(4)}\,+\,\sqrt{1\,-\gamma^2}\,\gamma^{(3)})\,\hat{\epsilon}_2\,,
\end{equation}
\begin{equation} \label{proj2}
\hat{\epsilon}_2\,=\,-\,(\mp\,\gamma\,\gamma^{(4)}\,+\,\sqrt{1\,-\gamma^2}\,\gamma^{(3)})\,\hat{\epsilon}_1\,.
\end{equation}
For the flat domain wall limit, $l\,\rightarrow\,\infty$, it reduced to the projection condition in \cite{Freedman:1999gp}.

Before proceeding to the spin-1/2 field variation, let us consider the projection condition in \eqref{proj1} and \eqref{proj2}. By multiplying $\gamma^{(4)}$ on both sides of \eqref{proj1} and \eqref{proj2} and rearranging them,
\begin{equation}
\gamma^{(4)}\,\left(
\begin{array}{l}
 \hat{\epsilon}_1 \\
 \hat{\epsilon}_2
\end{array}
\right)=i\,\left[\pm\,\gamma\,\left(
\begin{array}{ll}
 0 & -i \\
 i & \,\,\,\,0
\end{array}
\right)\,\,\left(
\begin{array}{l}
 \hat{\epsilon}_1 \\
 \hat{\epsilon}_2
\end{array}
\right)\,+\sqrt{1-\gamma^2}\,\left(
\begin{array}{l}
 -\,i\,\gamma^{(4)}\,\gamma^{(3)}\,\hat{\epsilon}_2 \\
 +\,i\,\gamma^{(4)}\,\gamma^{(3)}\,\hat{\epsilon}_1
\end{array}
\right)\right]\,.
\end{equation}
Without losing generality, we define an operator, $\mathbb{P}=-\,i\,\gamma^{(4)}\,\gamma^{(3)}$, which acts on the five dimensional spinors as
\begin{equation}
\mathbb{P}\,\hat{\epsilon}_1=-\cos\theta\,\hat{\epsilon}_1\,+\,\sin\theta\,\hat{\epsilon}_2\,,
\end{equation}
\begin{equation}
\mathbb{P}\,\hat{\epsilon}_2=+\sin\theta\,\hat{\epsilon}_1\,+\,\cos\theta\,\hat{\epsilon}_2\,,
\end{equation}
where $\theta=\theta(r)$. It satisfies the property, $\mathbb{P}^2$ = 1. Then the final projection condition is given by
\begin{equation} \label{fproj}
\gamma^{(4)}\,\hat{\epsilon}_i=i\,\left[\pm\,\gamma\,(\sigma^2)_{ij}\,+\,\sqrt{1-\gamma^2}\,\left(\cos\theta\,(\sigma^1)_{ij}\,+\,\sin\theta\,(\sigma^3)_{ij}\right)\right]\,\hat{\epsilon}_j\,,
\end{equation}
where $\sigma^i$, $i\,=\,1,\,2,\,3$, are the Pauli matrices. Similar projection condition was obtained in $\mathcal{N}$ = 2 gauged supergravity in \cite{LopesCardoso:2001rt, LopesCardoso:2002ec}.

Now we solve the spin-1/2 field variation. The variation  \eqref{s2} reduces to
\begin{equation} \label{s2r}
\hat{\epsilon}_i\,-\,m_{ij}\,\gamma^{(4)}\,\hat{\epsilon}_j\,=\,0\,,
\end{equation}
where
\begin{equation}
m_{ij}\,=\,\left(
\begin{array}{ll}
 m_1\,+\,m_2\,-\,m_3 & \,\,\,\,\,\,\,\,\,\, m_4\,-\,m_5 \\
 \,\,\,\,\,\,\,\,\,\, m_4\,+\,m_5 & -m_1\,+\,m_2\,-\,m_3
\end{array}
\right)\,,
\end{equation}
and
\begin{align}
m_1\,=&\,-\frac{2}{3\,g}\,i\,\text{csch}\chi\,\Big(\sin(a\,-\,\psi)\,\phi'\,+\,\frac{1}{2}\,\sinh(2\,\phi)\,\cos(a\,-\,\psi)\,a'\Big)\,, \notag \\
m_2\,=&\,-\frac{2}{3\,g}\,\sinh^2\phi\,a'\,, \notag \\
m_3\,=&\,-\frac{2}{3\,g}\,\psi'\,, \notag \\
m_4\,=&\,+\frac{2}{3\,g}\,i\,\text{csch}\chi\,\Big(\cos(a\,-\,\psi)\,\phi'\,+\,\frac{1}{2}\,\sinh(2\,\phi)\,\sin(a\,-\,\psi)\,a'\Big)\,, \notag \\
m_5\,=&\,+\frac{4}{3\,g}\,\text{csch}(2\,\chi)\,\chi'\,.
\end{align}
Using the projection condition \eqref{fproj} in \eqref{s2r}, we obtain the supersymmetry equations,
\begin{align} \label{su3se1}
\phi'\,=&\,+\,\frac{3}{2}\,g\,\sqrt{1-\gamma^2}\,\cos(a\,-\psi\,+\,\theta)\,\sinh\chi\,, \\ \label{su3se2}
\chi'\,=&\,\mp\,\frac{3}{4}\,g\,\gamma\,\sinh(2\,\chi)\,=\,\pm\,\frac{g}{2}\,\frac{\partial{W}}{\partial{\chi}}\,\gamma, \\ \label{su3se3}
a'\,=&\,-\,3\,g\,\sqrt{1-\gamma^2}\,\sin(a\,-\,\psi\,+\,\theta)\,\text{csch}(2\,\phi)\,\sinh\chi\,, \\ \label{su3se4}
\psi'\,=&\,+\,\frac{3}{2}\,g\,\sqrt{1-\gamma^2}\,\sin(a\,-\,\psi\,+\,\theta)\,\tanh\phi\,\sinh\chi\,.
\end{align}
We have also obtained the field equations, and they are presented in appendix E. We verified that the supersymmetry equations, \eqref{su3U} and \eqref{su3se1}-\eqref{su3se4}, are indeed consistent with the field equations, provided that one adds a condition on the phase of the projection, $\theta$, 
\begin{equation} \label{su3se5}
\theta'\,=\,-\,\frac{3}{2}\,g\,\sqrt{1-\gamma^2}\,\sin(a\,-\,\psi\,+\,\theta)\,\tanh\phi\,\sinh\chi\,.
\end{equation}
{\it Note that the supersymmetry equations imply that in the limit, $l\,\rightarrow\,\infty$, which describes a flat domain wall, we must set $\phi,\,a,\,\psi$ to be constants, {\it i.e.} the dilaton/axion fields decouple, and vice versa.} One can turn on the dilaton/axion fields only in the curved domain wall \cite{LopesCardoso:2001rt, LopesCardoso:2002ec}.

We have also checked the integrability of the spin-3/2 field variations for the $r$- and $z$-directions, but they do not generate any new constraint on the supersymmetry. The variations for these directions are presented in appendix C. On the other hand, by solving the spin-3/2 field variation for the $r$-direction, we obtain the $r$-dependence of the spinors \cite{Pilch:2011},
\begin{equation}
\left(
\begin{array}{l}
 \hat{\epsilon}_1(r) \\
 \hat{\epsilon}_2(r)
\end{array}
\right)\,=\,e^{U/2}\,
\left(
\begin{array}{ll}
 \,\,\,\,\,\cos\frac{\theta}{2} & \sin\frac{\theta}{2} \\
 -\sin\frac{\theta}{2} & \cos\frac{\theta}{2}
\end{array}
\right)
\left(
\begin{array}{ll}
 e^{\frac{i}{2}\,\Lambda} & \,\,\,\,0 \\
 \,\,\,\,0 & e^{-\frac{i}{2}\,\Lambda}
\end{array}
\right)
\left(
\begin{array}{l}
 \hat{\epsilon}_1^{(0)} \\
 \hat{\epsilon}_2^{(0)}
\end{array}
\right),
\end{equation}
where $\hat{\epsilon}_i^{(0)}$, $i$ = 1, 2, are independent of the $r$-coordinate, and $\cos\Lambda\,=\,\gamma$. This explains the fact that all the integrability conditions have been satisifed, $\it i.e.$ an explicit solution to a system of equations must satisfy all integrabilities automatically \cite{Pilch:2011}.

Let us also comment that one can show that the projection condition, \eqref{fproj}, is in fact identical to the spin-1/2 field variation, \eqref{s2r}, by using the supersymmetry equations, \eqref{su3U} and \eqref{su3se1}-\eqref{su3se4}. This proves that \eqref{fproj} is indeed the most general projection condition.

Now we numerically solve the supersymmetry equations, \eqref{su3U} and \eqref{su3se1}-\eqref{su3se4}. We choose the upper sign for $r\,>\,0$ and the lower sign for $r\,<\,0$ \cite{Clark:2005te}. There is a narrow range of initial conditions which gives smooth and nonsingular solutions,
\begin{equation} \label{range}
\frac{1}{4}\,g\,<\,e^{-U(0)}\,<\,\frac{3}{5}\,g\,,
\end{equation}
where the minimum is obtained from $0\,<\,\gamma\,<\,1$, and the maximum is from $U''\,>\,0$ \cite{Clark:2005te}. Outside of this range the solution becomes singular at the domain wall {\it i.e.} at the origin. A numerical solution in the critical range is plotted in figure 1, with the choice of initial conditions, $U(0)\,=\,0$, $\chi(0)\,=\,0.01$, $\psi(0)\,=\,0.1$, $\phi(0)\,=\,1$, $a(0)\,=\,0.1$, $\theta(0)\,=\,0.1$, and $g\,=\,2$. {\it Note that the five-dimensional dilaton and axion fields, $\phi$ and $a$, exhibit the dilaton profile of Janus solutions, {\it i.e.} it takes constant values on both sides of the interface, but jumps across the interface.} Indeed we will explicitly identify the solution to be the supersymmetric Janus solution in five dimensions in the next section.

\begin{figure}[h!]
\begin{center}
\includegraphics[width=2.0in]{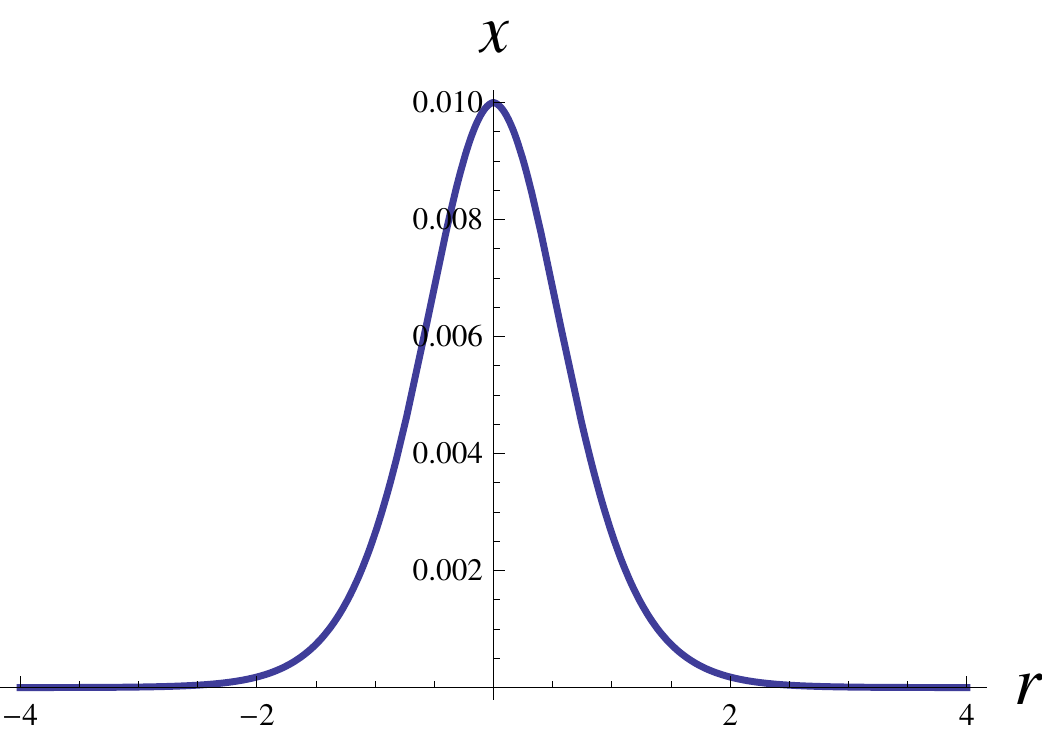} \qquad \includegraphics[width=2.0in]{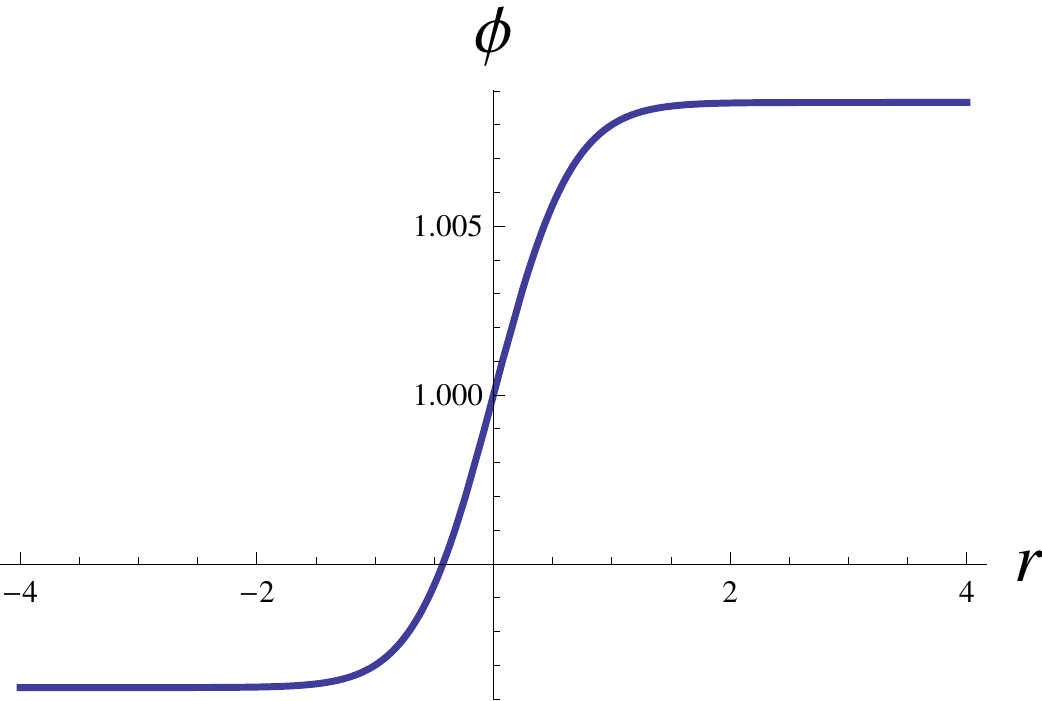} \qquad \includegraphics[width=2.0in]{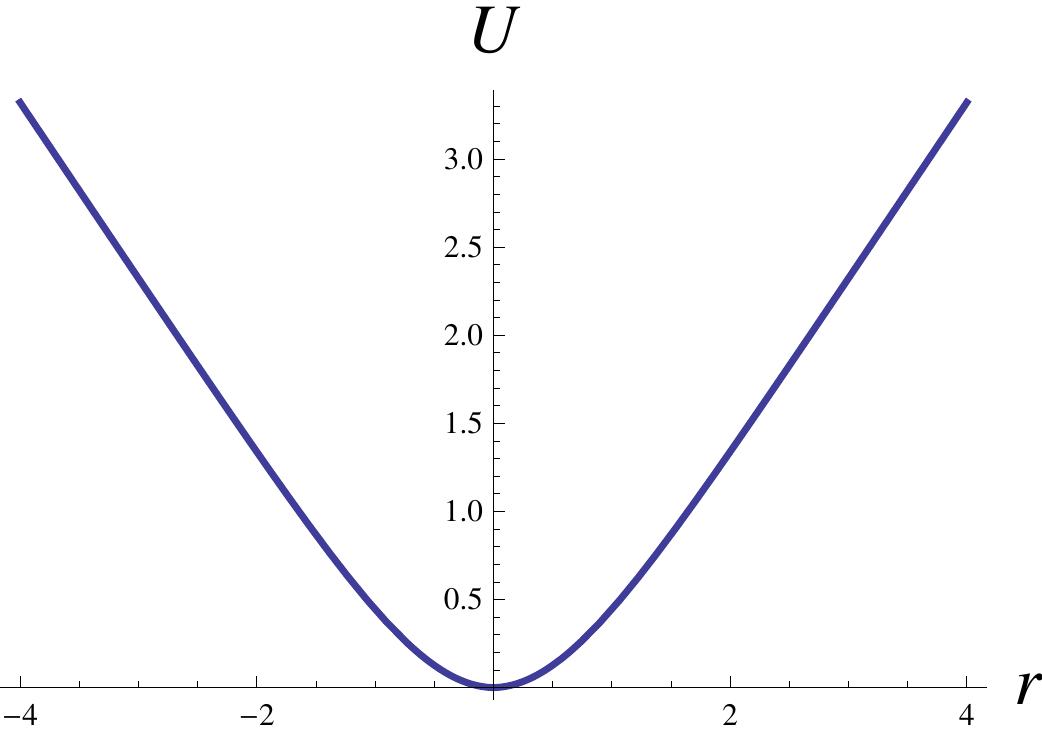}
\end{center}
\end{figure}
\begin{figure}[h!]
\begin{center}
\includegraphics[width=2.0in]{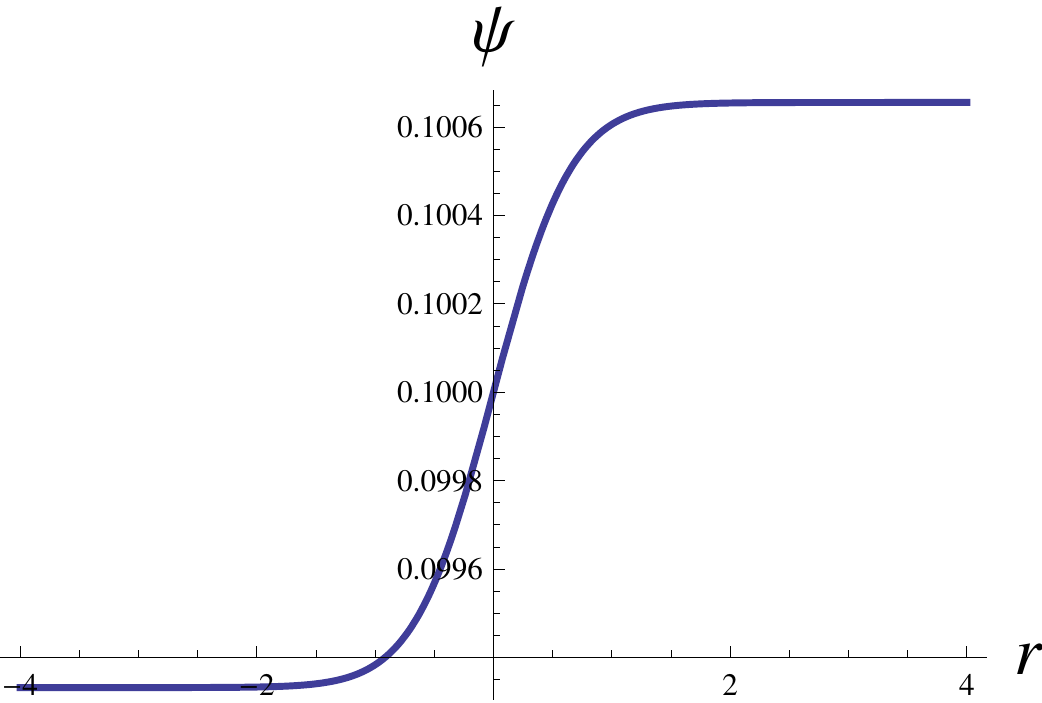} \qquad \includegraphics[width=2.0in]{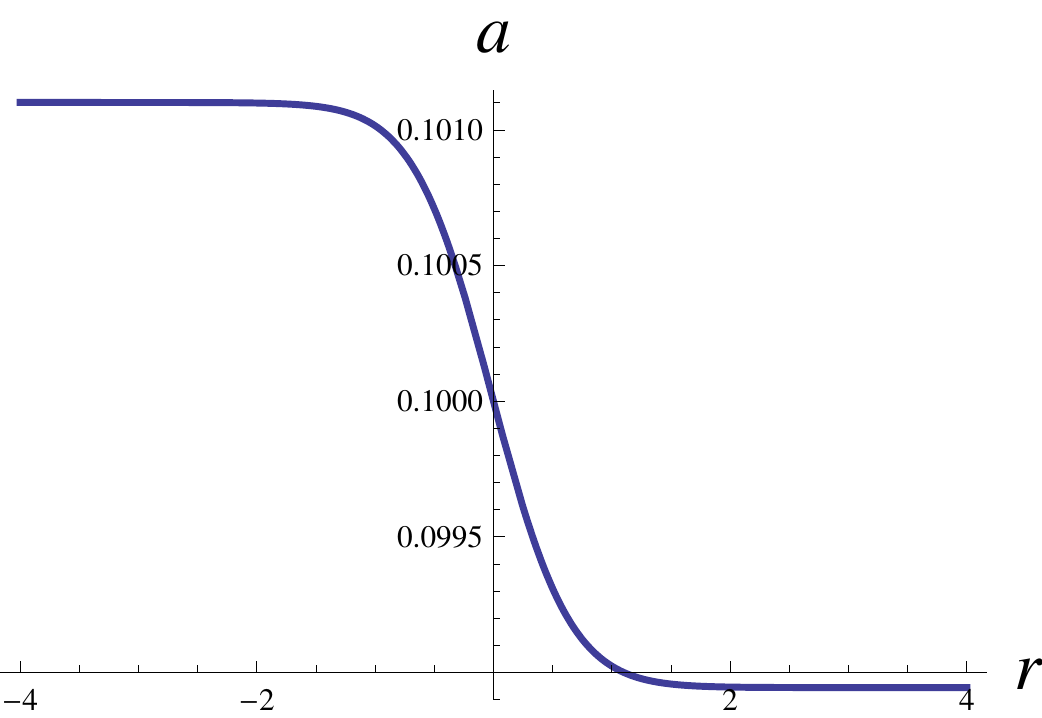} \qquad \includegraphics[width=2.0in]{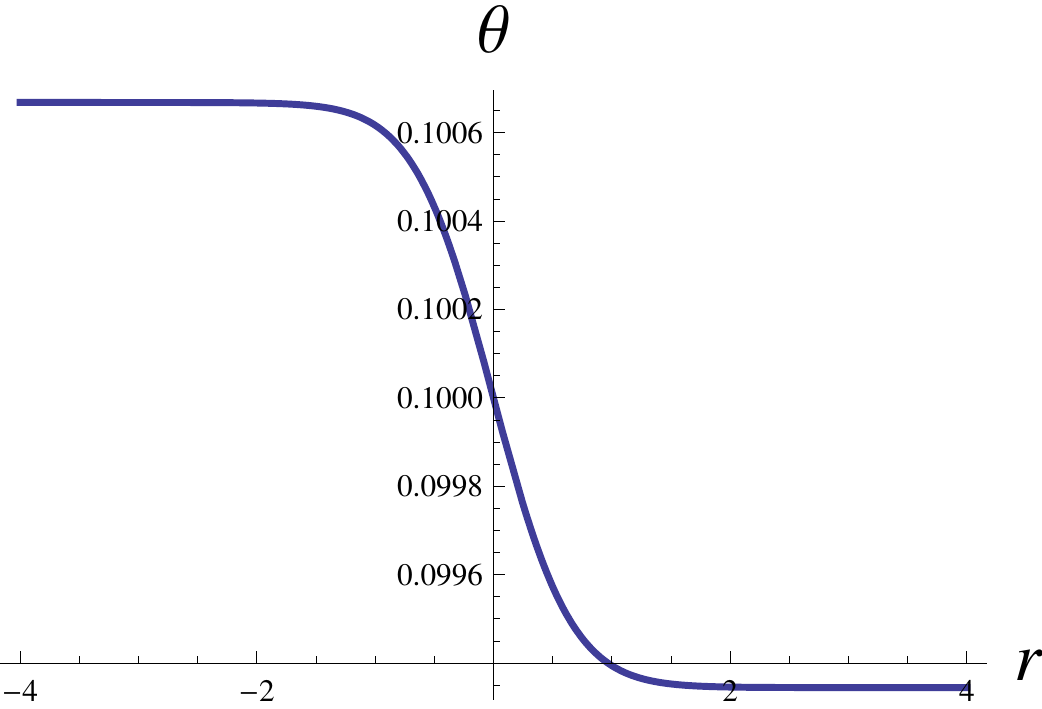}
\caption{{\it A numerical solution of the supersymmetry equations}}
\label{1}
\end{center}
\end{figure}

Before we close this section, let us count the number of supersymmetries the solution has. Each five-dimensional spinor, $\hat\epsilon_i$, $i\,=\,1,\,2$, has four complex components, so we have sixteen real supercharges in total to begin with. The Majorana-Weyl condition on five-dimensional spinors halves the number of real supercharges to eight. Then the only projection condition we have, \eqref{fproj}, also halves the number to four, so we have an $\mathcal{N}$ = 1 supersymmetric solution.

\section{Super Janus in $\mathcal{N}$ = 2 gauged supergravity in five dimensions \cite{Clark:2005te}}

There is a supersymmetric Janus solution, the super Janus, discovered by Clark and Karch in $\mathcal{N}$ = 2 gauged supergravity in five dimensions \cite{Clark:2005te}. In this section we will show that the solution in the $SU(3)$-invariant truncation in the previous section is indeed identical to the super Janus.

We briefly review $\mathcal{N}$ = 2 gauged supergravity with one hypermultiplet in five dimensions \cite{Ceresole:2000jd, Ceresole:2001wi}. The bosonic sector of the theory has a graviton $e_\mu\,^a$, a vector field $A_\mu$, and four scalar fields $q^X$. The scalar fields parametrize the coset manifold $\frac{SU(2,1)}{SU(2){\times}U(1)}$. The bosonic part of the Lagrangian is
\begin{equation} \label{sjlag}
e^{-1}\,\mathcal{L}\,=\,-\frac{1}{2}\,R\,-\,\frac{1}{2}\,g_{XY}\,D_{\mu}\,q^X\,D^{\mu}\,q^Y\,-\,\mathcal{P}(q)\,-\,\frac{1}{4}\,F_{\mu\nu}\,F^{\mu\nu}\,+\,\mathcal{L}_{CS}\,,
\end{equation}
where
\begin{equation}
D_{\mu}q^X\,=\,\partial_{\mu}q^X\,+\,g\,A_{\mu}\,K^X(q)\,,
\end{equation}
and $K^{X}$ are the Killing vectors of the gauged isometries on the scalar manifold. {\footnote { In this section, $g$ denotes the coupling constant of the super Janus, which is three times smaller than the one for the $SU(3)$-invariant truncation, $g^{\text{super Janus}}\,=\,\frac{1}{3}\,g^{SU(3)}$.}} Parametrizing the scalar fields by $q^X=\{V,\,\sigma,\,R,\,\alpha\}$, {\footnote { The scalar field, $R$, was denoted by $r$ in \cite{Clark:2005te}. It should not be confused with the Ricci scalar in \eqref{sjlag}.}} the scalar potential is given by
\begin{equation}
\mathcal{P}\,=\,g^2\,\left(-6\,-\,\frac{3R^2}{V}\,+\,\frac{3R^4}{V^2}\right)\,,
\end{equation}
and the superpotential is 
\begin{equation}
W\,=\,1\,+\,\frac{R^2}{V}.
\end{equation}
The metric $g_{XY}$ of the scalar manifold is 
\begin{equation}
ds^2\,=\,\frac{1}{2V^2}\,dV^2\,+\,\frac{1}{2V^2}\,d\sigma^2\,-\,\frac{2R^2}{V^2}\,{d\sigma}\,{d\alpha}\,+\,\frac{2}{V}\,{dR^2}\,+\,\frac{2R^2}{V}\,(1\,+\,\frac{R^2}{V})\,{d\alpha^2}\,.
\end{equation}
As $\mathcal{N}$ = 2 gauged supergravity with one hypermultiplet is equivalent to the $SU(3)$-invariant  truncation of $\mathcal{N}$ = 8 gauged supergravity, they have identical field content and the scalar manifold.

Now we briefly review the super Janus in $\mathcal{N}$ = 2 gauged supergravity \cite{Clark:2005te}. 
The metric is the $AdS$-domain wall, \eqref{AdSmet}, which we employed for the background of the $SU(3)$-invariant truncation in section 2. There are also four scalar fields,
\begin{equation} \label{sjscalars}
V\,=\,V(r)\,,  \,\,\,\,\sigma\,=\,\sigma(r)\,,  \,\,\,\,R\,=\,R(r)\,,  \,\,\,\,\alpha\,=\,\alpha(r)\,,
\end{equation} 
which depend on the $r$-coordinate only. The gauge field, $A_\mu$, will be suppressed. Then from the supersymmetry variations, one obtains the supersymmetry equations,
\begin{align}
U'\,&=\,\mp\,g\,W\,\gamma\,,
\\ \label{sjse1}
V'\,&=\,6\,g\,\left(\mp\,R^2\,\gamma\,+\,R\,\sqrt{V}\,\sqrt{1-\gamma^2}\right)\,,
\\ \label{sjse2}
R'\,&=\,3\,g\,\left(\pm\,R\,\gamma\,+\,\frac{R^2}{\sqrt{V}}\,\sqrt{1-\gamma^2}\right)\,,
\end{align}
where
\begin{equation}
\gamma\,=\,\sqrt{1\,-\,\frac{\lambda^2\,e^{-2U}}{g^2\,W^2}}\,,
\end{equation}
and the scalar fields $\sigma$ and $\alpha$ are consistently set to be constant.
Then, numerically plotting $V=V(r)$, we find that it exhibits the nontrivial profile of the dilaton field in Janus solutions.

Now we prove the equivalence of the super Janus and the solution in the $SU(3)$-invariant truncation. One can reparametrize $\{V,\,\sigma,\,R,\,\alpha\}$ in terms of $\{\chi,\,\psi,\,\phi,\,a\}$ by using the inhomogeneous coordinates, $\zeta_i$, $i$ = 1, 2, on the scalar manifold as an intermediate parametrization. We present the details of the reparametrization in appendix D. By employing the reparametrization to the action of the $SU(3)$-invariant truncation, \eqref{su3lag}, we find that it precisely reduces to the action of the super Janus, \eqref{sjlag}. Then, as the supersymmetry equations, \eqref{sjse1} and \eqref{sjse2}, are for the special case of constant $\sigma$ and $\alpha$, they turn out to be the supersymmetry equations of the $SU(3)$-invariant truncation, \eqref{su3se1}-\eqref{su3se4}, with the constant phases, {\it i.e.} $\psi$ and $a$ are constant, or more specifically, $a\,-\,\psi\,+\theta\,=\,0$. This proves that the solution of the $SU(3)$-invariant truncation considered in section 2 is indeed equivalent to the super Janus.

\bigskip
\bigskip
\bigskip
\bigskip
\bigskip

\section{Lift of the $SU(3)$-invariant truncation to type IIB supergravity}

We uplift the $SU(3)$-invariant truncation in section 2 to type IIB supergravity by the consistent truncation ansatz{\footnote { Here the consistent truncation ansatz means type IIB supergravity fields expressed in terms of five-dimensional fields of $\mathcal{N}$ = 8 gauged supergravity. On the other hand, there are consistent truncations from different directions, {\it i.e.} five-dimensional actions obtained by truncating type IIB supergravity, {\it e.g.} \cite{Cassani:2010uw, Gauntlett:2010vu}. We will consider them in section 7.}}. The consistent truncation ansatz for metric and dilaton/axion fields were presented in \cite{Khavaev:1998fb, Pilch:2000ue, Pilch:2000fu}. By employing the ansatz, lift of the $SU(3)$-invariant truncation was performed in \cite{Pilch:2000fu}, however, the five-dimensional dilaton/axion fields were suppressed. In this section we will lift the five-dimensional dilaton/axion fields, and as a consequence, we will have nontrivial IIB dilaton/axion fields. We postpone the lift of fluxes to section 6.

\subsection{The metric}

The ten-dimensional metric is given by 
\begin{equation}
ds^2\,=\,\Omega^2\,ds^2_{1,4}\,+\,ds^2_5\,,
\end{equation}
where $ds^2_{1,4}$ is an arbitrary solution of $\mathcal{N}$ = 8 gauged supergravity in five dimensions. In order to have Janus solution we employ the $AdS$-domain wall metric, \eqref{AdSmet}. The consistent truncation ansatz for the inverse metric of internal space is given by \cite{Khavaev:1998fb, Pilch:2000ue, Pilch:2000fu}
\begin{equation}
\Delta^{-\frac{2}{3}}\,g^{pq}=\frac{1}{a^2}\,K^{IJp}\,K^{KLq}\,\widetilde{\mathcal{V}}_{IJab}\,\widetilde{\mathcal{V}}_{KLcd}\,\Omega^{ac}\,\Omega^{bd},
\end{equation}
where $K^{IJp}$ are Killing vectors on round $S^5$, $\Omega^{ab}$ is a $USp(8)$ symplectic form, $\Delta=\det^{1/2}(g_{mp}\hat{g}^{pq})$, and $\hat{g}^{pq}$ is the inverse of the round $S^5$ metric. The $\Delta$ is obtained by taking the determinant on both sides of the ansatz, and $\Omega^2=\Delta^{-\frac{2}{3}}$ is the warp factor.

To apply the consistent truncation ansatz, we first prepare the proper coordinates in which the $SU(3)$ isometry of internal space is manifest \cite{Pilch:2000fu}. In Cartesian coordinates, $y^I$, $I$ = 1, $\ldots$ , 6, on $\mathbb{R}^6$, we think of $S^5$ defined by the surface $\Sigma_I\,(y^I)^2\,=\,1$. Let us introduce complex coordinates corresponding to the complex structure, $J_{IJ}$,
\begin{equation}
u^1\,=\,y^1\,+\,i\,y^2, \,\,\,\,\,\,\,\,\,\, u^2\,=\,y^5\,+\,i\,y^6, \,\,\,\,\,\,\,\,\,\, u^3\,=\,y^3\,+\,i\,y^4\,.
\end{equation}
We then introduce the complex coordinates where $\zeta^i$, $i\,=\,1,\, 2,$ are the complex projective coordinates on $\mathbb{CP}_2$, and $\varphi$ is the $U(1)$ Hopf fiber angle \cite{Pilch:2000fu},
\begin{equation}
\left(
\begin{array}{l}
 u^1 \\
 u^2
\end{array}
\right)\,=\,u^3\,\left(
\begin{array}{l}
 \zeta^1 \\
 \zeta^2
\end{array}
\right)\,, \,\,\,\,\,\,\,\,\,\,\, u^3\,=\,(1\,+\,\zeta^1\,\overline{\zeta}_1\,+\,\zeta^2\,\overline{\zeta}_2)^{-1/2}\,e^{-i\,\varphi}\,.
\end{equation}
Convenient real coordinates for the complex coordinates are \cite{Pilch:2000fu}
\begin{equation}
\left(
\begin{array}{l}
 \text{$\zeta^{1}$} \\
 \text{$\zeta^{2}$}
\end{array}
\right)\,=\,-\tan\theta\,\,g\,(\alpha_1,\,\alpha_2,\,\alpha_3)\, \left(
\begin{array}{l}
 \text{1} \\
 \text{0}
\end{array}
\right)\,,
\end{equation}
where g ($\alpha_1$,$\alpha_2$,$\alpha_3$) is an $SU(2)$ invariant matrix in terms of Euler angles, $\it{e.g.}$
\begin{equation}
g\,(\alpha_1,\,\alpha_2,\,\alpha_3)\,=\,\left(\begin{array}{ll}
 \,\,\,\,e^{-\frac{i}{2}\,(\alpha_1\,+\,\alpha_3)} \cos \left(\frac{\alpha_2}{2}\right) & \,\,\,\,e^{-\frac{i}{2}\,(\alpha_1\,-\,\alpha_3)} \sin
   \left(\frac{\alpha_2}{2}\right) \\
 -e^{+\frac{i}{2}\,(\alpha_1\,-\,\alpha_3)} \sin \left(\frac{\alpha_2}{2}\right) & \,\,\,\,e^{+\frac{i}{2}\,(\alpha_1\,+\,\alpha_3)} \cos
   \left(\frac{\alpha_2}{2}\right)
\end{array}
\right)\,.
\end{equation}

With the choice of above coordinates, the lifted metric of internal space reduces to
\begin{equation} \label{liftmet}
ds^2_5\,=\,\frac{1}{\cosh\chi}\,ds^2_{CP_2}\,+\,\cosh\chi\,(d\varphi\,+\,\frac{1}{2}\,\sin^2\theta\,\sigma_3)^2\,,
\end{equation}
where
\begin{equation} \label{intmet}
ds^2_{CP_2}\,=\,d\theta^2\,+\,\frac{1}{4}\,\sin^2\theta\,(\sigma^2_1\,+\,\sigma^2_2\,+\,\cos^2\theta\,\sigma^2_3)\,,
\end{equation}
\begin{equation}
\Omega\,=\,\cosh^{1/2}\,\chi\,,
\end{equation}
and $\sigma_{i}$ are the left-invariant one-forms of $SU(2)$,
\begin{align}
\sigma_1\,=&\,-\sin\alpha_2\,\cos\alpha_3\,d\alpha_1\,+\sin\alpha_3\,d\alpha_2\,, \notag \\
\sigma_2\,=&\,+\sin\alpha_2\,\sin\alpha_3\,d\alpha_1\,+\cos\alpha_3\,d\alpha_2\,, \notag \\
\sigma_3\,=&\,-\cos\alpha_2\,d\alpha_1\,-d\alpha_3\,,
\end{align}
which satisfy $d\sigma_i\,=\,\frac{1}{2}\,\epsilon_{ijk}\,\sigma_j\,\wedge\,\sigma_k$. As mentioned before, lift of the $SU(3)$-invariant truncation was performed in section 9 of \cite{Pilch:2000fu} without the five-dimensional dilaton/axion fields. Compared to the parametrization of internal space in \cite{Pilch:2000fu}, here we have $\alpha_i\rightarrow-\alpha_i\,,\theta\rightarrow-\theta\,,\varphi\rightarrow-\varphi$. Besides the parametrization, the lifted metric, \eqref{liftmet}, is identical to the one in \cite{Pilch:2000fu}, {\it i.e.} it is independent of the five-dimensional dilaton/axion fields, $\phi$ and $a$.

\subsection{The dilaton/axion fields}

The IIB dilaton/axion fields $(\Phi,\,C_{(0)})$ form a complex scalar, $\tau$, and are related to $B$ by
\begin{equation}
\tau\,=\,C_{(0)}\,+\,i\,e^{-\Phi}\,=\,i\,\frac{1-B}{1+B}\,,
\end{equation}
and $f$ is defined by
\begin{equation}
f\,=\,\frac{1}{\sqrt{1\,-\,|B|^2}}\,.
\end{equation}

The consistent truncation ansatz for the dilaton/axion fields is given by \cite{Pilch:2000ue}
\begin{equation} 
\Delta^{-\frac{4}{3}}\,(SS^T)^{\alpha\beta}\,=\,const\,\times\,\epsilon^{\alpha\gamma}\,\epsilon^{\beta\delta}\,\mathcal{V}_{I\gamma}\,\,^{ab}\,\mathcal{V}_{J\delta}\,\,^{cd}\,y^I\,y^J\,\Omega_{ac}\,\Omega_{bd}\,.
\end{equation}
From the ansatz the dilaton/axion field matrix, $S$, in the $SL(2,\mathbb{R})$ basis reduces to
\begin{equation} \label{Smat}
S\,=\,\frac{1}{2\sqrt{1-|B|^2}}\left(
\begin{array}{ll}
 2+(B+B^*) & i (B-B^*) \\
 i (B-B^*) & 2-(B+B^*)
\end{array}
\right),
\end{equation}
where
\begin{equation} \label{BB}
B\,=\,i\,e^{i\,a}\,\tanh\phi\,.
\end{equation}
By changing the basis to $SU(1,1)$, we obtain the dilaton/axion field matrix, $V$, \cite{Pilch:2000ue},
\begin{equation}
V\,=\,U^{-1}\,S\,U\,=\,f\,\left(
\begin{array}{ll}
 1 & B \\
 B^* & 1
\end{array}
\right)\,, \,\,\,\,\,\,\,\,\,\,\,\,\,\,\,\,\,\,\,\,\,\,\,\,\,\,\,\,\,\,
U\,=\,\left(
\begin{array}{ll}
 1 & \,\,\,1 \\
 i & -i
\end{array}
\right)\,,
\end{equation}
where
\begin{equation} \label{ff}
f\,=\,\cosh\phi\,.
\end{equation}
The IIB dilaton and axion fields are
\begin{align}\label{iibphi}
\Phi\,&=\,\ln\Big(\cosh(2\,\phi)\,-\,\sin(a)\,\sinh(2\,\phi)\Big)\,, \\ \label{iibc0}
C_{(0)}\,&=\,\frac{1}{\,\text{sec}(a)\,\coth(2\,\phi)\,-\,\tan(a)}\,,
\end{align}
and we note that they manifestly depend on the five-dimensional dilaton/axion fields, $\phi$ and $a$. In fugure 2 the IIB dilaton and axion fields are plotted with the identical initial condition as figure 1.
\begin{figure}[h!]
\begin{center}
\includegraphics[width=2.0in]{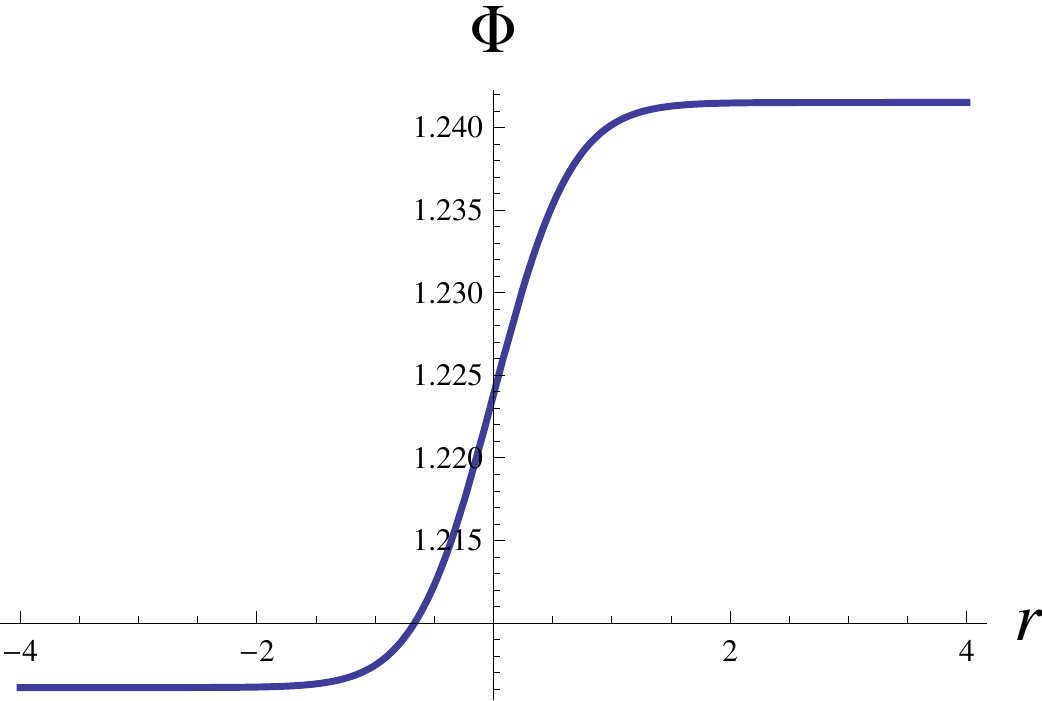} \qquad \includegraphics[width=2.0in]{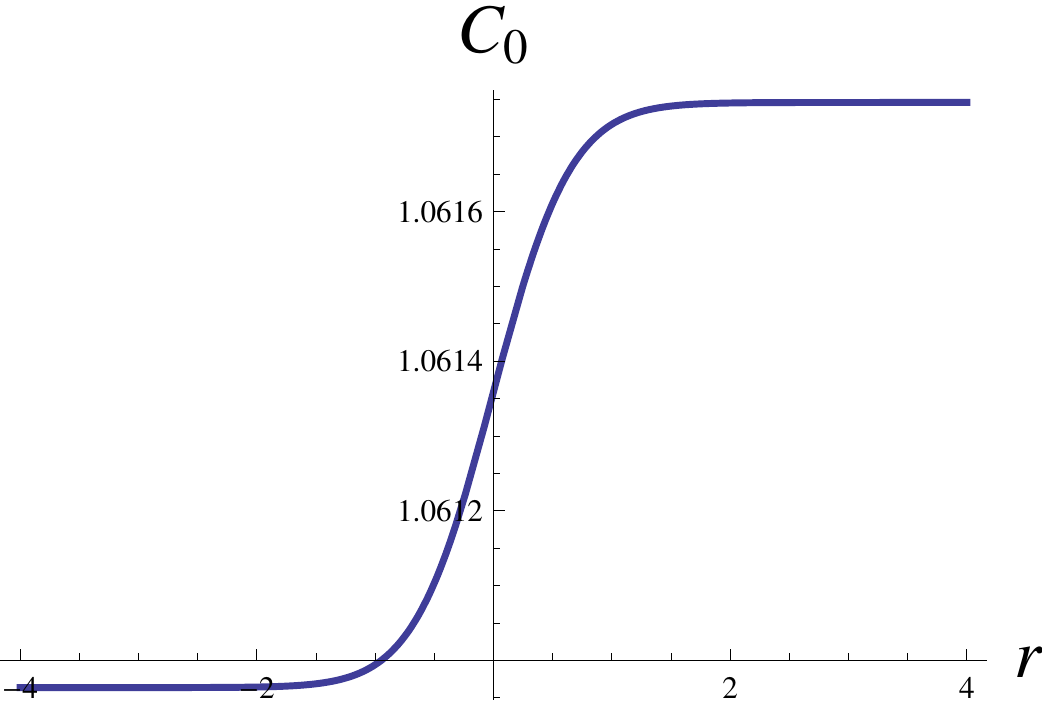}
\caption{{\it A numerical solution for the dilaton and axion fields}}
\label{1}
\end{center}
\end{figure}
Note that the dilaton and axion fields exhibit the dilaton profile of Janus solutions. Indeed we will explicitly identify our lifted solution as a special case of the supersymmetric Janus solution in type IIB supergravity in the next section.

\section{Supersymmetric Janus solution in type IIB supergravity \cite{D'Hoker:2006uu}}

As remarked in the introduction, the supersymmetric Janus solutions in type IIB supergravity were constructed by D'Hoker, Estes and Gutperle in \cite{D'Hoker:2006uu, D'Hoker:2007xy, D'Hoker:2007xz} with variety of supersymmetries and isometries. In this section we will show that by choosing metric and dilaton/axion fields to be the lifted ones in section 4, the supersymmetric Janus solution with $SU(3)$ isometry in \cite{D'Hoker:2006uu} is completely determined, {\it i.e.} it fixes all the IIB fields uniquely including three- and five-form fluxes.

We briefly review the supersymmetric Janus solution with the internal space isometry $SU(3)$ in type IIB supergravity \cite{D'Hoker:2006uu}. The metric is given by
\begin{equation} \label{jmet}
ds^2\,=\,f_4^2\,(d\mu^2\,+\,ds^2_{AdS_4})\,+\,f_1^2\,(d\beta\,+\,A_1)\,+\,f_2^2\,ds^2_{CP_2}\,,
\end{equation}
where 
\begin{equation}
A_1\,=\,\frac{1}{2}\,\sin^2\theta\,\sigma_3\,,
\end{equation}
and $\sigma_i,$ $i$ = 1, 2, 3, are the $SU(2)$-invariant one-forms.
The five-form flux is given by
\begin{equation} \label{jfiveform}
F_{(5)}\,=\,f_5\,\big(-e^0\,{\wedge}\,e^1\,{\wedge}\,e^2\,{\wedge}\,e^3\,{\wedge}\,e^4+e^5\,{\wedge}\,e^6\,{\wedge}\,e^7\,{\wedge}\,e^8\,{\wedge}\,e^9\big)\,,
\end{equation}
where $e^n$, $n\,=\,0,\,\ldots,\,9$ are the frames of the metric.
The two-form gauge potential is given by
\begin{equation} \label{jthreeform}
B_{(2)}^{\text{there}}\,=\,C_{(2)}\,-\,i\,B_{(2)}\,=\,i\,f_3\,\Omega_2\,-\,i\,\overline{g}_3\,\overline{\Omega}_2\,,
\end{equation}
where $C_{(2)}$ and $B_{(2)}$ are RR and NSNS two-form gauge potentials respectively, $\Omega_2$ is the holomorphic (2,0)-form on $S^5$, $f_3$ and $g_3$ are complex functions, and the bar denotes complex conjugation. The dilaton/axion fields are denoted by $B$ with its associated function $f$. Overall, the most general solution with the $SU(3)$ isometry of internal space is specified by the seven functions, $f_1$, $f_2$, $f_3$, $g_3$, $f_4$, $f_5,$ and $B$, and they depend only on the $\mu$-coordinate.

In section 9 of \cite{D'Hoker:2006uu}, a special case is presented when
\begin{equation}
a\,=\,-\,\frac{3}{f_1\,f_2^2}\,f\,(f_3\,-\,B\,g_3)\,=\,0\,,
\end{equation}
where $a$ is a function defined for convenience in \cite{D'Hoker:2006uu}.
Furthermore, in this case, 
\begin{equation}
f_1\,f_2\,=\,\rho\,, \,\,\,\,\,\,\,\,\,\, f_5\,=\,\frac{3}{2\,f_1}\,-\,\frac{1}{2}\,\frac{f_1}{f_2^2}\,,
\end{equation}
where $\rho$ is a constant, and some functions are integrated to hyper-elliptic integral as
\begin{equation}
\left(\frac{\partial\Psi}{\partial{\mu}}\right)^2\,=\,\left(1\,+\,\frac{C_2^2}{9\rho^8}\,\Psi^6\right)^2\,-\,\Psi^2\,,
\end{equation}
where $\Psi\,=\,\psi^{\text{there}}\,=\,\frac{\rho}{f_2f_4}$ and $C_2$ is a constant.

Now we compare the lifted metric, \eqref{liftmet}, and the dilaton/axion fields, \eqref{BB} and \eqref{ff}, in section 4 with the supersymmetric Janus solution presented here. By comparing the metric and the dilaton/axion fields, we find that the functions are given by 
\begin{equation}
f_1\,=\,\cosh^{1/2}\chi\,,
\end{equation}
\begin{equation}
f_2\,=\,\cosh^{-1/2}\chi\,,
\end{equation}
\begin{equation}
f_4\,=\,e^{U}\,\cosh^{1/2}\chi\,,
\end{equation}
\begin{equation}
B\,=\,i\,e^{i\,a}\,\tanh\phi\,, \,\,\,\,\,\,\,\,\,\,\,\,\,\,\,
f\,=\,\cosh\phi\,.
\end{equation}
Then we can plug the above set of functions into the field equations, (6.6), (6.13)-(6.16), and supersymmetry equations, (7.24)-(7.29), in \cite{D'Hoker:2006uu} to solve for the rest of functions, and we obtain 
\begin{equation} \label{f5}
f_5\,=\,-\frac{\cosh(2\,\chi)-5}{4\cosh^{1/2}\chi}\,,
\end{equation}
\begin{equation} \label{f3}
f_3\,=\,e^{i\,(a\,-\,\psi)}\,\sinh\phi\,\tanh\chi\,,
\end{equation}
\begin{equation} \label{g3}
g_3\,=\,-i\,e^{-i\,\psi}\,\cosh\phi\,\tanh\chi\,.
\end{equation}
These functions uniquely fixes three- and five-form fluxes in \eqref{jfiveform} and \eqref{jthreeform}. Furthermore, we note that this choice of the functions falls into the special case, $a\,=\,0$, explained above, and we obtain the hyper-elliptic integral,
\begin{equation}
\left(\frac{\partial{U}}{\partial{\mu}}\right)^2\,=\,e^{2\,U}\,+\,\frac{2}{9}\,C_2^2\,e^{-4U}\,+\,\frac{1}{4}\,\left(\frac{2}{9}\,C_2^2\right)^2\,e^{-10\,U}\,-\,1\,,
\end{equation}
where $\Psi\,=\,e^{U}$ and $\rho\,=\,1$. This proves that the lifted metric and the dilaton/axion fields from the $SU(3)$-invariant truncation in section 4 indeed gives a special case of the supersymmetric Janus solution in type IIB supergravity in \cite{D'Hoker:2006uu}.

\bigskip
\bigskip
\bigskip
\bigskip
\bigskip
\bigskip

\section{Lift of the $SU(3)$-invariant truncation to type IIB supergravity (continued)}

In this section we continue the lift of the $SU(3)$-invariant truncation to type IIB supergravity. We uplift the three- and five-form fluxes which were not lifted in section 4. The lift formulae for three- and five-form fluxes were proposed in \cite{Khavaev:2001yg}, however, we will find that the formulae do not work for the curved domain walls. We will propose modified lift formulae for three- and five-form fluxes valid for both the flat and the curved domain walls, and check them for some nontrivial cases including the $SU(3)$-invariant truncation.

\subsection {The three-form flux}

The three-form flux is defined by, {\it e.g.} \cite{Cassani:2010uw, Gauntlett:2010vu},
\begin{align}
G_{(3)}\,&=\,dC_{(2)}\,-\,\tau\,dB_{(2)} \notag \\
\,&=\,dC_{(2)}\,-\,(C_{(0)}\,+\,i\,e^{-\Phi})\,dB_{(2)} \notag \\
\,&=\,(dC_{(2)}\,-C_{(0)}\,dB_{(2)})\,-\,i\,e^{-\Phi}\,dB_{(2)} \notag \\
\,&=\,F_{(3)}\,-\,i\,e^{-\Phi}\,H_{(3)}\,,
\end{align}
where $C_{(2)}$ and $B_{(2)}$ are RR and NSNS two-form gauge potentials respectively, and we also define 
\begin{equation}
F_{(3)}\,=\,dC_{(2)}\,-\,C_{(0)}\,dB_{(2)}\,,
\end{equation}
\begin{equation}
H_{(3)}\,=\,dB_{(2)}\,.
\end{equation}

The lift formula for the two-form gauge potential was proposed in \cite{Khavaev:2001yg}, 
\begin{equation} \label{3f}
B^\alpha\,_{pq}\,=\,k\,L^2\,\mathcal{M}^{\alpha\beta}\,(y^K\,\mathcal{V}_{K\alpha}\,^{ab})\,\left(\mathcal{V}_{IJab}\,\frac{\partial{y}^I}{\partial\xi^p}\,\frac{\partial{y}^J}{\partial\xi^q}\right)\,,
\end{equation}
where $y^I$ are the Cartesian coordinates for an $\mathbb{R}^6$ embedding of $S^5$, $\xi^p$ are the intrinsic coordinates on the $S^5$, $\mathcal{M}\,=\,S\,S^T$, and $S$ is given in \eqref{Smat}. However, if we apply the formula to the $SU(3)$-invariant truncation with dilaton and axion fields, it does not produce the correct two-form gauge potential found in section 5, \eqref{jthreeform} with \eqref{f3} and \eqref{g3}. By empirical observation we propose a modified lift formula for two-form gauge potential, 
\begin{equation} \label{3f}
B_{\alpha\,pq}\,=\,-\frac{i}{\sqrt{2}}\,\Delta^{-\frac{4}{3}}\,(y^K\,\mathcal{V}_{K\alpha}\,^{ab})\,\left(\mathcal{V}_{IJab}\,\frac{\partial{y}^I}{\partial\xi^p}\,\frac{\partial{y}^J}{\partial\xi^q}\right)\,,
\end{equation}
where $\Delta$ is the warp factor, and $B_1\,=\,B_{(2)}$, $B_2\,=\,C_{(2)}$. We have verified that this lift formula indeed produces the correct two-form gauge potential in section 5. There is also another combination of two-form gauge potentials,
\begin{equation} \label{threeform}
A_{(2)}\,=\,C_{(2)}\,-\,\tau\,B_{(2)}\,=\,\frac{e^{i\psi}\,\tanh\chi}{\cosh\phi\,+\,i\,e^{ia}\sinh\phi}\,\overline{\Omega}_2\,,
\end{equation}
where
\begin{equation}
\Omega_2\,=\,\frac{1}{12}\,e^{-3\,i\,\varphi}\,\sin\theta\left(2\,i\,d\theta\,\wedge\,(\sigma_1\,+\,i\,\sigma_2)\,+\,\frac{1}{2}\,\sin(2\,\theta)\,(\sigma_1\,+\,i\,\sigma_2)\,\wedge\,\sigma_3\right)\,,
\end{equation}
is the holomorphic (2,0)-form of the internal space \cite{Pilch:2000fu}.

\subsection {The five-form flux}

The lift formula for five-form flux was also proposed in \cite{Khavaev:2001yg}, 
however, we found that it does not produce the correct five-form flux for the $SU(3)$-invariant truncation with dilaton/axion fields in section 5. In this subsection we propose a modified lift formula for five-form flux from empirical observations. 

We consider the metric,
\begin{equation}
ds^2_{1,4}\,=\,e^{2\,U(\mu)}\,ds^2_4\,+\,d\mu^2\,,
\end{equation}
where $ds^2_{1,4}$ is any solution of $\mathcal{N}$ = 8 gauged supergravity in five-dimensions, and $vol_5$ denotes the unit volume form of $ds^2_{1,4}$, and $vol_4$ of $ds^2_4$. We define the geometric $W$-tensors,
\begin{align}
\widetilde{W}_{ab}\,=&\,-\,\epsilon^{\alpha\beta}\,y^I\,y^J\,\Omega^{cd}\,\mathcal{V}_{I\alpha{ac}}\,\mathcal{V}_{J\beta{bd}}\,, \\
\widetilde{W}_{abcd}\,=&\,+\,\epsilon^{\alpha\beta}\,y^I\,y^J\,\mathcal{V}_{I\alpha{ab}}\,\mathcal{V}_{J\beta{cd}}\,,
\end{align}
and the geometric scalar potential,
\begin{equation}
\mathcal{\widetilde{P}}\,=\,-\,\frac{g^2}{32}\,\big(2\,W_{ab}\,\widetilde{W}^{ab}\,-\,W_{abcd}\,\widetilde{W}^{abcd}\,\big)\,.
\end{equation}
The geometric superpotential, $\widetilde{W}$, is one of the eigenvalues of $\widetilde{W}_{ab}$.

Before presenting the modified lift formula, let us review the lift formula proposed in \cite{Khavaev:2001yg},
\begin{equation} 
F_{(5)}\,=\,\mathcal{F}\,+\,*\mathcal{F}\,,
\end{equation}
where
\begin{equation}
\mathcal{F}\,=\,d\,(\widetilde{W}\,vol_4)\,.
\end{equation}
Applying this formula to the $SU(3)$-invariant truncation with dilaton and axion fields, we obtain
\begin{equation}
\mathcal{F}\,=\,\frac{32}{g^2}\,\mathcal{P}\,\gamma\,vol_5=\,\cosh^2\chi\,\big(\cosh(2\,\chi)\,-\,5\big)\,\gamma\,vol_5\,,
\end{equation}
where $\mathcal{P}$ is the scalar potential and $\gamma$ is from the supersymmetry equations invoked when taking the derivative of the geometric superpotential. However, it is not the correct five-form flux, \eqref{jfiveform} with \eqref{f5}, as the correct one does not have the factor of $\gamma$.{\footnote { In fact the five-form flux in \eqref{jfiveform} with \eqref{f5}, is also the correct five-form flux in \cite{Pilch:2000fu}, which is the lift of the $SU(3)$-invariant truncation in the flat domain wall {\it i.e.} without dilaton and axion fields. Hence, for the flat domain wall, $\gamma\,=\,1$, and the formula produces the correct five-form flux in \cite{Pilch:2000fu}.}}

Now we propose the modified lift formula for five-form flux,
\begin{equation} \label{5f}
\mathcal{F}\,=\,\frac{32}{g^2}\,\widetilde{\mathcal{P}}\,vol_5\,+\,\frac{\partial\widetilde{W}}{\partial\xi^p}\,d\xi^p\,\wedge\,vol_4,
\end{equation}
where $\xi^p$ are the intrinsic coordinates of internal space.

By employing the lift formula to the $SU(3)$-invariant truncation with dilaton and axion fields, we obtain that $\widetilde{\mathcal{P}}\,=\,\mathcal{P}$, $\widetilde{W}\,=\,W$, so $\frac{\partial\widetilde{W}}{\partial\xi^p}\,=\,0$. Hence, the five-form flux is
\begin{equation}
F_{(5)}\,=\,\cosh^2\chi\,\big(\cosh(2\,\chi)\,-\,5\big)\,vol_5\,-\,\frac{\cosh(2\,\chi)\,-\,5}{2\,\cosh^2\chi}\,J_2\,\wedge\,J_2\,\wedge(\eta\,+\,A)\,,
\end{equation}
where $J_2$ is the K\"ahler form, and $\eta$ is the one-form dual to the Reeb Killing vector to be explained more in section 7. This is indeed the five-form flux found in section 5. We believe that the modified lift formula, \eqref{5f}, generates the correct five-form fluxes for all the flat domain wall cases that the lift formula in \cite{Khavaev:2001yg} was tested. So far we have verified that it does produce the correct five-form flux for the $SU(2){\times}U(1)$-invariant truncation in section 3 of \cite{Pilch:2000fu}.

However, the lift formula only gives the terms of five-form flux which do not involve the gauge field, $A_\mu$, in five dimensions. For the complete five-form flux, we will just present the flux obtained by using the results in \cite{Cassani:2010uw, Gauntlett:2010vu},
\begin{align}
F_{(5)}\,=\,&\cosh^2\chi\big(\cosh(2\,\chi)\,-\,5\big)\,vol_5\,-\,\frac{1}{2}\,*\mathcal{K}\,\wedge\,(\eta\,+\,A)\,-\,*(dA)\,\wedge\,J_2 \notag\\  -&\frac{\cosh(2\,\chi)\,-\,5}{2\cosh^2\chi}\,J_2\,\wedge\,J_2\,\wedge\,(\eta\,+\,A)\,-\,\frac{1}{4\cosh^4\chi}\,\mathcal{K}\,\wedge\,J_2\,\wedge\,J_2\,-\,dA\,\wedge\,J_2\,\wedge\,(\eta\,+\,A)\,,
\end{align}
where
\begin{equation}
\mathcal{K}\,=\,-\,\sinh^2(2\,\chi)\,\left(\partial_\mu\psi\,+\sinh^2\phi\,\partial_\mu{a}\,+\,g\,A_\mu\right)\,dx^\mu\,.
\end{equation}

In this section we proposed the lift formulae for three- and five-form fluxes, \eqref{3f} and \eqref{5f}. However, we should stress that we have not derived them from a consistent truncation of type IIB supergravity, but have constructed them based on empirical observations. It is possible that some modification would be needed in the general case, as they are modifications of the formulae in \cite{Khavaev:2001yg}.

\section {Type IIB supergravity on Sasaki-Einstein manifolds \cite{Cassani:2010uw, Gauntlett:2010vu}}

Recently there has been notable development in consistent truncation of type IIB supergravity on Sasaki-Einstein manifolds \cite{Cassani:2010uw, Gauntlett:2010vu, Liu:2010sa, Skenderis:2010vz}. In this section, we will show that the $SU(3)$-invariant truncation of $\mathcal{N}$ = 8 gauged supergravity in five dimensions and its lift to type IIB supergravity in section 2, 4 and 6 provide a particular example of consistent truncation in \cite{Cassani:2010uw, Gauntlett:2010vu}.

Locally the Sasaki-Einstein metric can be written as \cite{Cassani:2010uw, Gauntlett:2010vu}
\begin{equation}
ds^2\,(SE_5)\,=\,ds^2\,(KE_4)\,+\,\eta\,\otimes\,\eta\,,
\end{equation}
where $ds^2\,(KE_4)$ is a local K\"ahler-Einstein metric with positive curvature and $\eta$ is a globally defined one-form dual to the Reeb Killing vector. There are also a globally defined K\"ahler two-form $J_2$ and a (2, 0)-form complex structure $\Omega_2$, and they satisfy 
\begin{align}
d\eta\,=\,&2\,J_2\,, \\
d\Omega_2\,=\,&3\,i\,\eta\,\wedge\,\Omega_2\,.
\end{align} 
The type IIB metric is then given by \cite{Cassani:2010uw, Gauntlett:2010vu}
\begin{equation}
ds^2\,=\,e^{\frac{2}{3}\,(4U\,+\,V)}\,ds_{(E)}^2\,+\,e^{2U}\,ds^2\,(KE_4)\,+\,e^{2V}\,(\eta\,+\,A)\,\otimes\,(\eta\,+\,A)\,,
\end{equation}
where $ds_{(E)}^2$ is an arbitrary metric on an external five-dimensional spacetime, $U$ and $V$ are scalar functions {\footnote { Here $U$ and $V$ have nothing to do with the warp factor, $U$, in \eqref{AdSmet} and the scalar field, $V$, in \eqref{sjscalars}.}} and $A$ is a one-form defined on the external five-dimensional spacetime.

In \cite{Cassani:2010uw, Gauntlett:2010vu} it was shown that the consistent truncation of type IIB supergravity on Sasaki-Einstein manifolds leads to $\mathcal{N}$ = 4 gauged supergravity coupled to two vector mulptiplets in five dimensions. In section 5.3 and 5.4 of \cite{Cassani:2010uw} and section 3.4.8 of \cite{Gauntlett:2010vu}, a particular truncation is presented, and for instance, the five-dimensional action for the particular truncation is {\footnote { In the truncation in section 3.4.8 of \cite{Gauntlett:2010vu}, the dilaton/axion fields were not considered.}}
\begin{align} \label{SEkin}
\mathcal{L}_{kin}\,=&\,-\,\frac{1}{2}\,\partial_\mu\sigma\,\partial^\mu\sigma\,-\,\frac{1}{8}\,\sinh^2(2\sigma)\,(\partial_\mu\theta\,-\,\frac{1}{2}\,e^{\Phi}\,\partial_\mu{C_{(0)}}\,-\,3\,A_\mu)^2\, \notag \\ &-\,\frac{1}{8}\,\cosh^2\sigma\,(\partial_\mu\Phi\,\partial^\mu\Phi\,+\,e^{2\,\Phi}\,\partial_\mu{C_{(0)}}\,\partial^\mu{C_{(0)}})\,, \\
\mathcal{P}\,=&\,+\,\frac{3}{32}\,g^2\,\left(\cosh^2(2\sigma)\,-\,4\,\cosh(2\sigma)\,-\,5\,\right)\,,
\end{align}
where $\sigma$ and $\theta$ are five-dimensional scalar fields, {\footnote { Here $\sigma$ and $\theta$ have nothing to do with the scalar field, $\sigma$, in \eqref{sjscalars} and the phase, $\theta$ in \eqref{fproj}.}} and $\Phi$ and $C_{(0)}$ are dilaton and axion fields of type IIB supergravity respectively. This truncation without the dilaton/axion fields was used to construct a holographic superconductor in \cite{Gubser:2009qm, Gubser:2009gp}.

We found that the following reparametrization of the particular truncation precisely reproduces the five-dimensional action, \eqref{su3lag}, and the lifted IIB fields of the $SU(3)$-invariant truncation in section 4 and 6, {\footnote { We refer to appendix D for derivation of this reparametrization}}
\begin{align} \label{su3toSE}
\sigma\,&=\,\chi\,, \notag \\
\theta\,&=\,\text{Tan}^{-1}\,\left(\frac{\cos\psi\,-\,\sin(a\,-\,\psi)\,\tanh\phi}{\sin\psi\,-\,\cos(a\,-\,\psi)\,\tanh\phi}\right)\,, \notag \\
\Phi\,&=\,\ln\Big(\cosh(2\,\phi)\,-\,\sin(a)\,\sinh(2\,\phi)\Big)\,, \notag \\
C_{(0)}\,&=\,\frac{1}{\,\text{sec}(a)\,\coth(2\,\phi)\,-\,\tan(a)}\,.
\end{align}
This proves that the $SU(3)$-invariant truncation of $\mathcal{N}$ = 8 gauged supergravity and its lift indeed provides a particular example of type IIB supergravity on Sasaki-Einstein manifolds in \cite{Cassani:2010uw, Gauntlett:2010vu}.

\section{Conclusions}

In this paper, we studied the $SU(3)$-invariant truncation of $\mathcal{N}$ = 8 gauged supergravity in five dimensions with dilaton and axion fields and its lift to type IIB supergravity. Furthermore, we showed that the two known supersymmetric Janus solutions in five and in ten dimensions, {\it i.e.} the super Janus in five dimensions \cite{Clark:2005te} and the supersymmetric Janus solution with $SU(3)$ isometry in type IIB supergravity \cite{D'Hoker:2006uu}, are constructed in a unified way in the framework of $\mathcal{N}$ = 8 gauged supergravity and its lift.

As an application of the method presented here, one would construct the supersymmetric Janus solution with $SU(2){\times}U(1)$ isometry in type IIB supergravity. As mentioned in the introduction, according to the classification of Janus solutions in type IIB supergravity \cite{D'Hoker:2006uv}, there are four kinds of solutions with $SO(6)$, $SU(3)$, $SU(2){\times}U(1)$ and $SO(3){\times}SO(3)$ isometries, and each of them has zero, four, eight, and sixteen Poincar\'e supersymmetries, respectively. However, the supersymmetric Janus solution with isometry of $SU(2){\times}U(1)$ has not been constructed explicitly so far. On the other hand, there have been detailed studies on the $SU(2){\times}U(1)$-invariant truncation of $\mathcal{N}$ = 8 gauged supergravity in five dimensions \cite{Khavaev:1998fb, Freedman:1999gp} and its lift to type IIB supergravity \cite{Pilch:2000fu}, but they consider the flat domain wall solutions without dilaton and axion fields. However, its $\mathcal{N}$ = 2 gauged supergravity counterpart was studied on the curved domain wall with dilaton and axion fields \cite{LopesCardoso:2002ff}. It would be interesting to study the $SU(2){\times}U(1)$-invariant truncation of $\mathcal{N}$ = 8 gauged supergravity with dilaton and axion fields and its lift to type IIB supergravity. With this truncation, we would be able to construct the supersymmetric Janus solution with $SU(2){\times}U(1)$ isometry in five and ten dimensions.

As a different direction of generalization, one would consider the $SU(3)$-invariant truncation with the dilaton and axion fields, and also with the gauge fields. Recently, the $SU(2){\times}U(1)$-invariant truncation \cite{Khavaev:1998fb, Pilch:2000fu, Khavaev:2001yg} was revisited in \cite{Bobev:2010de} with the five-dimensional gauge fields dual to chemical potential in the boundary field theory. The domain wall solution then describes the RG flow interpolating two global $AdS_5$ spaces with chemical potentials \cite{Bobev:2010de}. The $SU(3)$-invariant truncation with dilaton and axion fields, also with the gauge fields would have supersymmetric Janus solutions with chemical potential, and it would be interesting to study this from the AdS/CMT perspective. Especially, as no black hole solutions have been constructed in the Janus geometries, it would be interesting to see if there are charged black hole solutions in the $SU(3)$-invariant truncation.

\bigskip
\medskip
\leftline{\bf Acknowledgements}
It is a pleasure to thank Krzysztof Pilch for suggesting the problem and his help throughout this project. This work was supported in part by the DOE grant DE-FG03-84ER-40168.

\appendix
\section{$\mathcal{N}$ = 8 gauged supergravity in five dimensions}
\renewcommand{\theequation}{A.\arabic{equation}}
\setcounter{equation}{0} 

In this appendix we review $\mathcal{N}$ = 8 gauged supergravity in five dimensions with emphasis on the structure of its scalar manifold, $E_{6(6)}/USp(8)$, by following \cite{Gunaydin:1985cu}. We will employ the conventions of \cite{Gunaydin:1985cu} throughout the paper.

The $SO(6)$ gauged $\mathcal{N}$ = 8 supergravity in five dimensions \cite{Pernici:1985ju, Gunaydin:1984qu, Gunaydin:1985cu} has local $USp(8)$ symmetry, but global $E_{6(6)}$ symmetry of the ungauged theory is broken. The field content consists of 1 graviton $e_{\mu}\,^{a}$, 8 gravitini $\psi_{\mu}\,^{a}$, 15 vector fields $A_{{\mu}IJ}$, 12 two-form tensor fields $B_{\mu\nu}\,^{I\alpha}$, 48 spinor fields $\chi^{abc}$, and 42 scalar fields $\phi^{abcd}$ where $a$, $b$, $\ldots$ are $USp(8)$ indices, $I$, $J$, $\ldots$ are $SL(6,\mathbb{R})$, and $\alpha$, $\beta$, $\ldots$ are $SL(2,\mathbb{R})$. Here $SL(6,\mathbb{R})$${\times}$$SL(2,\mathbb{R})$ is one of the maximal subgroups of $E_{6(6)}$.  

The infinitesimal $E_{6(6)}$ transformation in the $SL(6,\mathbb{R})$${\times}$$SL(2,\mathbb{R})$ basis, ($z_{IJ}$, $z^{I{\alpha}}$), in terms of $\Lambda^I\,_J$, $\Lambda^\alpha\,_\beta$, and $\Sigma_{IJK{\alpha}}$ was already given in \eqref{zls1} and \eqref{zls2}. Exponentiating the transformation in \eqref{zls1} and \eqref{zls2},
\begin{align} \label{zU1}
z'_{IJ}\,=\,&\,\,\frac{1}{2}\,U^{MN}\,_{IJ}\,z_{MN}\,+\,\sqrt{\frac{1}{2}}\,U_{P{\beta}IJ}\,z^{P{\beta}}\,, \\ \label{zU2}
z'^{K{\beta}}\,=\,&\,\,U_{P{\beta}}\,^{K{\alpha}}\,z^{P{\beta}}\,+\,\sqrt{\frac{1}{2}}\,U^{IJK{\alpha}}\,z_{IJ}\,,
\end{align}
we obtain the coset representatives in the $SL(6,\mathbb{R})$${\times}$$SL(2,\mathbb{R})$ basis, $U^{IJ}\,_{KL}$, $U^{IJK{\alpha}}$ and $U_{I{\alpha}}\,^{J{\beta}}$. We also have the coset representatives in the $USp(8)$ basis,
\begin{align} \label{UV1}
\mathcal{V}^{IJab}\,=\,&\,\,\frac{1}{8}\,\,\left[(\Gamma_{KL})^{ab}\,U^{IJ}\,_{KL}\,+\,2(\Gamma_{K{\beta}})^{ab}\,U^{IJK{\beta}}\right]\,, \\ \label{UV2}
\mathcal{V}_{I{\alpha}}\,^{ab}\,=\,&\,\,\frac{1}{4}\,\sqrt{\frac{1}{2}}\,\,\left[(\Gamma_{KL})^{ab}\,U_{I\alpha}\,^{KL}\,+\,2(\Gamma_{K{\beta}})^{ab}\,U_{I\alpha}\,^{K{\beta}}\right]\,.
\end{align}
The inverse coset representatives are
\begin{align}
\widetilde{\mathcal{V}}_{IJab}\,&=\,\frac{1}{8}\,[(\Gamma_{KL})_{ab}\,\widetilde{U}_{IJ}\,^{KL}\,+\,2\,(\Gamma_{K\alpha})_{ab}\,\widetilde{U}_{IJ}\,^{K\alpha}]\,, \\
\widetilde{\mathcal{V}}^{I\alpha}\,_{ab}\,&=\,\frac{1}{4}\,\sqrt{\frac{1}{2}}\,[(\Gamma_{KL})_{ab}\,\widetilde{U}^{I{\alpha}KL}\,+\,2\,(\Gamma_{K\beta})_{ab}\,\widetilde{U}_{I\alpha}\,^{K\beta}]\,.
\end{align}

Now we consider the action of the theory \cite{Gunaydin:1985cu}. The bosonic part of the Lagrangian is
\begin{equation} \label{N8lag}
e^{-1}\,\mathcal{L}\,=\,-\frac{1}{4}\,R\,+\,\mathcal{L}_{kin}\,+\,\mathcal{P}\,-\frac{1}{8}\,H_{{\mu}{\nu}{a}{b}}\,H^{{\mu}{\nu}{a}{b}}\,+\,\frac{1}{8\,g\,e}\,\epsilon^{\mu\nu\rho\sigma\tau}\,\epsilon_{\alpha\beta}\,B_{\mu\nu}\,^{I\alpha}\,D_{\rho}B_{\sigma\tau}\,^{I\beta}
\,+\,\mathcal{L}_{CS}\,,
\end{equation}
where the covariant derivative is defined by
\begin{equation}
D_{\mu}X_{aI}\,=\,\partial_{\mu}X_{aI}\,+\,Q_{{\mu}a}\,^{b}\,X_{bI}\,-\,g\,A_{{\mu}IJ}\,X_{aJ}\,,
\end{equation}
with the $USp(8)$ connection,
\begin{equation}
Q_{{\mu}a}\,^b\,=\,-\frac{1}{3}\,\left[\,\widetilde{\mathcal{V}}^{bcIJ}\,\partial_{\mu}\mathcal{V}_{IJac}\,+\,\widetilde{\mathcal{V}}^{bcI{\alpha}}\,{\partial}_{\mu}\mathcal{V}_{I{\alpha}ac}\,+\,g\,A_{{\mu}IL}\,\eta^{JL}\,(2\,V_{ae}\,^{IK}\,\widetilde{\mathcal{V}}^{be}\,_{JK}\,-\,\mathcal{V}_{J{\alpha}ae}\,\widetilde{\mathcal{V}}^{beI{\alpha}})\right]\,.
\end{equation}
The kinetic term for scalar fields is defined by
\begin{equation}
\mathcal{L}_{kin}\,=\,\,\frac{1}{24}\,P_{{\mu}abcd}\,P^{{\mu}abcd}\,,
\end{equation}
where
\begin{equation}
P_{\mu}\,^{abcd}\,=\,\widetilde{\mathcal{V}}^{ab}\,_{IJ}\,D_{\mu}\mathcal{V}^{IJcd}\,+\,\widetilde{\mathcal{V}}^{abI{\alpha}}\,D_{\mu}\mathcal{V}_{I{\alpha}}\,^{cd}\,.
\end{equation}
The scalar potential is defined by
\begin{equation}
\mathcal{P}\,=\,-\,\frac{1}{32}\,(2W_{ab}\,W^{ab}\,-\,W_{abcd}W^{abcd})\,,
\end{equation}
where
\begin{equation}
W_{abcd}\,=\,\epsilon^{\alpha\beta}\,\eta^{IJ}\,\mathcal{V}_{I{\alpha}ab}\,\mathcal{V}_{J{\beta}cd}\,,
\end{equation}
\begin{equation}
W_{ab}\,=\,W^c\,_{acb}\,.
\end{equation}
We also define
\begin{equation}
H_{\mu\nu}\,^{ab}\,=\,F_{\mu\nu}\,^{ab}+B_{\mu\nu}\,^{ab}\,, \end{equation}
where
\begin{align}
F_{\mu\nu}\,^{ab}\,&=\,F_{\mu\nu{IJ}}\,\mathcal{V}^{IJab}\,, \\
B_{\mu\nu}\,^{ab}\,&=\,B_{\mu\nu}\,^{I{\alpha}}\,\mathcal{V}_{I\alpha}\,^{ab}\,,
\end{align}
for the last three terms of Lagrangian.

We adopt the gamma matrix convention of \cite{Gunaydin:1985cu}, with
\begin{equation}
\{\gamma^{(i)},\,\gamma^{(j)}\}\,=\,2\,\eta^{ij}\,,
\end{equation}
where $\eta^{ij}\,=\,diag\,(+,\,-,\,-,\,-,\,-)$, and $\gamma^{(0)},\, \gamma^{(1)},\,\gamma^{(2)},\,\gamma^{(3)}$ are pure imaginary as in four-dimensions and $\gamma^{(4)}\,=\,i\,\gamma^{(5)}$ is pure real. The matrices $\gamma^{(0)}$ and $\gamma^{(5)}$ are antisymmetric and $\gamma^{(1)},\, \gamma^{(2)},\,\gamma^{(3)}$ are symmetric.

\section{$SU(2,1)$ algebra}
\renewcommand{\theequation}{B.\arabic{equation}}
\setcounter{equation}{0} 

$SU(2,1)$ algebra is given by
\begin{equation}
[L_i,\,L_j]\,=\,i\,f_{ijk}\,L_k\,,
\end{equation}
with the structure constants
\begin{equation}
f_{123}\,=\,1\,,\,\,\,\,f_{147}\,=\,f_{165}\,=f_{246}\,=\,f_{257}\,=\,f_{345}\,=\,f_{376}\,=\,\frac{1}{2}\,,\,\,\,\,f_{458}\,=\,f_{678}\,=\,-\frac{\sqrt{3}}{2}\,.
\end{equation}
The standard 3-dimensional $SU(2,1)$ generators are obtained by modifying $SU(3)$ Gell-Mann matrices where the Gell-Mann matrices are
\begin{multline}
\lambda_1\,=\,\left(
\begin{array}{lll}
 0 & 1 & 0 \\
 1 & 0 & 0 \\
 0 & 0 & 0
\end{array}
\right),\,\,\,\, \lambda_2\,=\,\left(
\begin{array}{lll}
 0 & -i & 0 \\
 i & 0 & 0 \\
 0 & 0 & 0
\end{array}
\right),\,\,\,\,\lambda_3\,=\,\left(
\begin{array}{lll}
 1 & 0 & 0 \\
 0 & -1 & 0 \\
 0 & 0 & 0
\end{array}
\right),\,\,\,\,\lambda_4\,=\,\left(
\begin{array}{lll}
 0 & 0 & 1 \\
 0 & 0 & 0 \\
 1 & 0 & 0
\end{array}
\right),\\\lambda_5\,=\,\left(
\begin{array}{lll}
 0 & 0 & -i \\
 0 & 0 & 0 \\
 i & 0 & 0
\end{array}
\right),\,\,\,\,\lambda_6\,=\,\left(
\begin{array}{lll}
 0 & 0 & 0 \\
 0 & 0 & 1 \\
 0 & 1 & 0
\end{array}
\right),\,\,\,\,\lambda_7\,=\,\left(
\begin{array}{lll}
 0 & 0 & 0 \\
 0 & 0 & -i \\
 0 & i & 0
\end{array}
\right),\,\,\,\,\lambda_8\,=\,\frac{1}{\sqrt{3}}\left(
\begin{array}{lll}
 1 & 0 & 0 \\
 0 & 1 & 0 \\
 0 & 0 & -2
\end{array}
\right)\,.
\end{multline} 
Multiplying four of the Gell-Mann matrices by $i$, they close onto an $SU(2,1)$ algebra,
\begin{equation}
L_1\,=\,\frac{\lambda_1}{2}\,,\,\,\,\, L_2\,=\,\frac{\lambda_2}{2}\,,\,\,\,\,L_3\,=\,\frac{\lambda_3}{2}\,,\,\,\,\,L_4\,=\,i\frac{\lambda_4}{2}\,,\,\,\,\,L_5\,=\,i\frac{\lambda_5}{2}\,,\,\,\,\,L_6\,=\,i\frac{\lambda_6}{2}\,,\,\,\,\,L_7\,=\,i\frac{\lambda_7}{2}\,,\,\,\,\,L_8\,=\,\frac{\lambda_8}{2}\,,
\end{equation} 
where $L_1$, $L_2$, $L_3$ are $SU(2)$ generators, $L_4$, $L_5$, $L_6$, $L_7$ are $\frac{SU(2,1)}{SU(2){\times}U(1)}$ coset generators, and $L_8$ is a $U(1)$ generator.

The generators in the 27-dimensional representation in section 2.2 corresponding to the 3-dimensional generators are given by
\begin{multline}
L_1\,{\rightarrow}\,\frac{i}{8}(\Sigma^{(3)}-\Sigma^{(4)}),\,\,\,\,
L_2\,{\rightarrow}\,\frac{i}{8}(\Sigma^{(3)}+\Sigma^{(4)}),\,\,\,\,
L_3\,{\rightarrow}\,\frac{i}{4}(\Lambda^{(5)}-\Lambda^{(8)}),\,\,\,\,
L_4\,{\rightarrow}\,\frac{i}{4\sqrt{2}}\Sigma^{(1)},\\
L_5\,{\rightarrow}\,\frac{i}{4\sqrt{2}}\Sigma^{(2)},\,\,\,\,
L_6\,{\rightarrow}\,\frac{i}{2\sqrt{2}}(\Lambda^{(7)}+\Lambda^{(6)}),\,\,\,\,
L_7\,{\rightarrow}\,\frac{i}{2\sqrt{2}}(\Lambda^{(7)}-\Lambda^{(6)}),\,\,\,\,
L_8\,{\rightarrow}\,\frac{i}{4\sqrt{3}}(\Lambda^{(5)}+3\Lambda^{(8)}).
\end{multline}

\section{The supersymmetry variations for spin-3/2 fields}
\renewcommand{\theequation}{C.\arabic{equation}}
\setcounter{equation}{0} 

In this appendix we present the $SU(3)$-invariant truncation of supersymmetry variations for spin-3/2 fields. The variation for $t$-, $x$-, $y$- directions is given in \eqref{vartxy}. For $z$-direction the variation is given by
\begin{align}
-2&\,e^{-U}\,\gamma^{(3)}\,z\,\partial_z\,\hat{\epsilon}_1\,-U'\,\gamma^{(4)}\,\hat{\epsilon}_1\,-\frac{1}{3}\,g\,W\,\hat{\epsilon}_2\,=\,0\,, \\
+2&\,e^{-U}\,\gamma^{(3)}\,z\,\partial_z\,\hat{\epsilon}_2\,+U'\,\gamma^{(4)}\,\hat{\epsilon}_2\,-\frac{1}{3}\,g\,W\,\hat{\epsilon}_1\,=\,0\,.
\end{align}
For the variation in the $r$-direction we need to know the action of $Q_{\mu{a}}\,^b$ tensor on the spinors,
\begin{align}
Q_{ra}\,^b\,\eta_{(1)b}\,&=\,+\,Q_1\,\eta_{(1)a}\,+\,Q_2\,\eta_{(2)a}\,, \\
Q_{ra}\,^b\,\eta_{(2)b}\,&=\,-\,\overline{Q}_2\,\eta_{(1)a}\,-\,Q_1\,\eta_{(2)a}\,,
\end{align}
where
\begin{align}
Q_1=&-i\,\sinh\chi\,\Big[\cos(a\,-\,\psi)\,\phi'\,-\,\frac{1}{2}\,\sin(a\,-\,\psi)\,\sinh(2\,\phi)\,a'\Big]\,, \\
Q_2=&-i\,\Big[\sinh\chi\,\left(\sin(a\,-\,\psi)\,\phi'\,-\,\frac{i}{2}\,\sinh\chi\,\psi'\right)\, \notag \\ &+\,\frac{1}{2}\,\left(\cos(a\,-\,\psi)\,\sinh(2\,\phi)\,\sinh\chi\,-\,\frac{i}{2}\,\big(-3\,+\,\cosh(2\,\chi)\big)\,\sinh^2\phi\,\right)\,a'\Big]\,.
\end{align}
Then the variation in the $r$-direction is given by
\begin{equation}
\partial_r\,\hat{\epsilon}_1\,-\,(+\,Q_1\,\hat{\epsilon}_1\,+\,Q_2\,\hat{\epsilon}_2)\,+\,\frac{1}{6}\,g\,W\,\gamma^{(4)}\,\hat{\epsilon}_2=0\,,
\end{equation}
\begin{equation}
\partial_r\,\hat{\epsilon}_2\,-\,(-\,\overline{Q}_2\,\hat{\epsilon}_1\,-\,Q_1\,\hat{\epsilon}_2)\,-\,\frac{1}{6}\,g\,W\,\gamma^{(4)}\,\hat{\epsilon}_1=0\,,
\end{equation}
where $W$ is the superpotential in \eqref{sp}.

\section{The parametrizations of the scalar manifold}
\renewcommand{\theequation}{D.\arabic{equation}}
\setcounter{equation}{0} 

In this paper we have employed several different parametrizations for the four real scalar fields living on the scalar manifold, $\frac{SU(2,1)}{SU(2){\times}U(1)}$. In this appendix we summarize the origins of and the relations between different parametrizations. 

The coset manifold, $\frac{SU(2,1)}{SU(2){\times}U(1)}$, is topologically an open ball in $\mathbb{C}^2$ with the Bergman metric \cite{BrittoPacumio:1999sn},
\begin{equation} \label{bergman}
ds^2\,=\,\frac{d\zeta_1\,d\overline\zeta_1\,+\,d\zeta_2\,d\overline\zeta_2}{1\,-\,\zeta_1\,\overline\zeta_1\,-\,\zeta_2\,\overline\zeta_2}\,+\,\frac{(\overline\zeta_1\,d\zeta_2\,+\,\overline\zeta_2\,d\zeta_2)(\zeta_1\,d\overline\zeta_2\,+\,\zeta_2\,d\overline\zeta_2)}{(1\,-\,\zeta_1\,\overline\zeta_1\,-\,\zeta_2\,\overline\zeta_2)^2}\,,
\end{equation}
which is a K\"ahler metric with K\"ahler potential,
\begin{equation}
\mathcal{K}\,=\,-\,\frac{1}{2}\,\ln(1\,-\,\zeta_1\,\overline\zeta_1\,-\,\zeta_2\,\overline\zeta_2)\,.
\end{equation}

The first two parametrizations of the scalar manifold we employed in this paper were the rectangular and angular parametrizations, $\{x_1,\,x_2,\,x_3,\,x_4\}$ in \eqref{exp} and $\{\chi,\,\psi,\,\phi,\,a\}$, respectively, for the the $SU(3)$-invariant truncation in section 2. The relation between them is given in \eqref{su3para}. In terms of the rectangular parametrization, the inhomogeneous coordinates on the scalar manifold are given by
\begin{align} \label{zx1}
\zeta_1\,=\,&\frac{(x_1\,+\,ix_2)\,\tanh\left(\frac{1}{2}\,\sqrt{x_1^2\,+\,x_2^2}\right)}{\sqrt{x_1^2\,+\,x_2^2}}\,\text{sech}\left(\frac{1}{2}\,\sqrt{x_3^2\,+x_4^2}\right)\,, \\ \label{zx2}
\zeta_2\,=\,&\frac{(x_3\,+\,ix_4)\,\tanh\left(\frac{1}{2}\,\sqrt{x_3^2\,+\,x_4^2}\right)}{\sqrt{x_3^2\,+\,x_4^2}}\,.
\end{align}
We can reverse the relation to get
\begin{equation} \label{xtoz1}
x_1\,=\,\frac{\zeta_1\,+\overline{\zeta}_1}{2\,Z_1}\,, \,\,\,\,\,\,\,\,\,\, x_2\,=\,\frac{\zeta_1\,-\overline{\zeta}_1}{2\,i\,\,Z_1}\,, \,\,\,\,\,\,\,\,\,\,
x_3\,=\,\frac{\zeta_2\,+\overline{\zeta}_2}{2\,Z_2}\,, \,\,\,\,\,\,\,\,\,\, x_4\,=\,\frac{\zeta_2\,-\overline{\zeta}_2}{2\,i\,\,Z_2}\,,
\end{equation}
where
\begin{equation} \label{xtoz2}
Z_1\,=\,\frac{\sqrt{\zeta_1\,\overline{\zeta}_1}\,\sqrt{1\,+\,\zeta_2\,\overline{\zeta}_2}}{2\,\text{Tanh}^{-1}\sqrt{\zeta_1\,\overline{\zeta_1}}}\,, \,\,\,\,\,\,\,\,\,\,\,\,\,\,\,\,\,\,\,\, Z_2\,=\,\frac{\sqrt{\zeta_2\,\overline{\zeta}_2}}{2\,\text{Tanh}^{-1}\sqrt{\zeta_2\,\overline{\zeta_2}}}\,.
\end{equation}

Before proceeding to the third parametrization, we consider the $SU(3)$-invariant truncation in terms of the complex coordinates, $\zeta_i$, $i$ = 1, 2. When we exponentiate the coset generators in \eqref{exp}, if we employ the complex coordinates by \eqref{xtoz1}, we can have the action of the $SU(3)$-invariant truncation in terms of the complex coordinates,
\begin{equation} \label{complag}
e^{-1}\,\mathcal{L}\,=\,-\,\frac{1}{4}\,R\,+\,\mathcal{L}_{kin}\,+\,\mathcal{P}\,-\,\frac{3}{4}\,F_{\mu\nu}\,F^{\mu\nu}\,+\,\mathcal{L}_{CS}\,.
\end{equation}
The kinetic term is
\begin{equation}
\mathcal{L}_{kin}\,=\,\frac{1}{2}\,h_{i\overline{j}}\,D_\mu\,\zeta_i\,D^\mu\,\overline{\zeta}_{\overline{j}}\,,
\end{equation}
where the metric is the Bergman metric, \eqref{bergman}, and the covariant derivative with respect to the gauge field is
\begin{equation}
D_\mu\,\zeta_1\,=\,\partial_\mu\,\zeta_1\,+\,3\,g\,A_\mu\,\zeta_1\,, \,\,\,\,\,\,\,\,\,\, D_\mu\,\zeta_2\,=\,\partial_\mu\,\zeta_2\,.
\end{equation}
The scalar potential is
\begin{equation}
\mathcal{P}\,=\,-\,\frac{3}{8}\,g^2\,\frac{(1\,-\,|\zeta_2|^2)\,(2\,-\,3|\zeta_1|^2\,-\,2|\zeta_2|^2)}{(1\,-\,|\zeta_1|^2\,-\,|\zeta_2|^2)^2}\,.
\end{equation}

Thirdly, in $\mathcal{N}$ = 2 gauged supergravity in five dimensions, there is another parametrization by the scalar fields, $\{V,\,\sigma,\,R,\,\alpha\}$, which was employed for the super Janus in section 3. In terms of these scalar fields the inhomogeneous coordinates are given by \cite{Ceresole:2001wi, Clark:2005te}
\begin{align} \label{zit1}
\zeta_{1}\,=\,&\frac{-\,2\,i\,R\,e^{i\,\alpha}}{1\,+\,R^2\,+\,V\,+\,i\, \sigma}\,, \\ \label{zit2}
\zeta_{2}\,=\,&\frac{1\,-\,R^2\,-\,V\,-\,i\,\sigma}{1\,+\,R^2\,+\,V\,+\,i\,\sigma}\,.
\end{align}
By plugging \eqref{zit1}, \eqref{zit2} into \eqref{complag}, we precisely reproduce the action of the super Janus, \eqref{sjlag}. The rest of the truncation, {\it e.g.} the supersymmetry equations, can also be reparametrized, and they are explained in section 3. This reparametrization was used to establish the equivalence of the $SU(3)$-invariant truncation and the super Janus in section 3.

Lastly, there is a parametrization by $\{\sigma,\, \theta,\, \Phi,\, C_{(0)}\}$, employed for a particular truncation of type IIB supergravity on Sasaki-Einstein manifolds in section 7. The $\Phi$ and $C_{(0)}$ are the IIB dilaton and axion fields respectively, and $\sigma$ and $\theta$ are some five-dimensional scalar fields. We briefly mention that by comparing Killing vectors for \eqref{su3kin} and \eqref{SEkin}, we have found the relation between $\{\sigma,\, \theta,\, \Phi,\, C_{(0)}\}$ and $\{\chi,\,\psi,\,\phi,\,a\}$ in \eqref{su3toSE}. Note that the IIB dilaton and axion fields are indeed identical to the ones from the lift in \eqref{iibphi} and \eqref{iibc0}. 

\section{The field equations of the $SU(3)$-invariant truncation}
\renewcommand{\theequation}{E.\arabic{equation}}
\setcounter{equation}{0} 

In this appendix, we present the field equations of the $SU(3)$-invariant truncation. Let us consider the action for complex scalar fields and gravity,
\begin{equation}
\mathcal{L}\,=\,\sqrt{g}\,\left(-\,\frac{1}{4}\,R\,+\,\frac{1}{2}\,g^{\mu\nu}\,h_{a\overline{b}}\,\partial_\mu\phi_a\,\partial_\nu\overline{\phi}_{\overline{b}}\,-\,\mathcal{P}(\phi_a,\overline{\phi}_{\overline{a}})\right)\,.
\end{equation}
The scalar equations reduce to
\begin{align}
\frac{1}{\sqrt{g}}\,\partial_\mu(\sqrt{g}\,g^{\mu\nu}\,\partial_\nu\phi^a)\,+\,\Gamma^a\,_{bc}\,g^{\mu\nu}\,\partial_\mu\phi^b\,\partial_\nu\phi^c\,-\,h^{\overline{b}a}\,\partial_{\overline{b}}\,\mathcal{P}&\,=\,0\,, \\
\frac{1}{\sqrt{g}}\,\partial_\mu(\sqrt{g}\,g^{\mu\nu}\,\partial_\nu\overline{\phi}^{\overline{a}})\,+\,\Gamma^{\overline{a}}\,_{\overline{b}\overline{c}}\,g^{\mu\nu}\,\partial_\mu\overline{\phi}^{\overline{b}}\,\partial_\nu\overline{\phi}^{\overline{c}}\,-\,h^{\overline{a}b}\,\partial_{b}\mathcal{P}&\,=\,0\,,
\end{align}
where
\begin{align}
\Gamma^{a}\,_{bc}&\,=\,h^{\overline{d}a}\,\partial_c{h}_{b\overline{d}}\,, \\
\Gamma^{\bar{a}}\,_{\overline{b}\bar{c}}&\,=\,h^{\overline{a}d}\,\partial_{\overline{c}}{h}_{d\overline{b}}\,.
\end{align}
The Einstein equations are
\begin{equation}
R_{\mu\nu}\,-\,\frac{1}{2}R\,g_{\mu\nu}\,=\,2\,T_{\mu\nu}\,,
\end{equation}
where the energy-momentum tensor is
\begin{equation}
T_{\mu\nu}\,=\,h_{a\overline{b}}\,\partial_\mu\phi_a\,\partial_\nu\overline{\phi}_{\overline{b}}\,-\,g_{\mu\nu}\left(\frac{1}{2}g^{\rho\sigma}h_{a\overline{b}}\partial_\rho\phi_a\partial_\sigma\overline{\phi}_{\overline{b}}\,-\,\mathcal{P}(\phi,\overline{\phi})\right)\,.
\end{equation}
For the metric in \eqref{ads4}, the Einstein equations reduce to
\begin{equation}
3\,(U''\,+\,2\,U'\,U')\,+\,\frac{3}{l^2}\,e^{-2U}\,=\,-\,2\,(\frac{1}{2}\,h_{a{\overline{b}}}\,\phi_a'\,\overline{\phi}_{\overline{b}}'\,-\,\mathcal{P})\,,
\end{equation}
\begin{equation}
3\,U'\,U'\,+\,\frac{3}{l^2}\,e^{-2U}\,=\,(\frac{1}{2}\,h_{a{\overline{b}}}\,\phi_a'\,\overline{\phi}_{\overline{b}}'\,+\,\mathcal{P})\,.
\end{equation}

Then, for the $SU(3)$-invariant truncation in \eqref{complag}, in terms of the inhomogeneous coordinates, $\{\zeta_1,\,\zeta_2\}$, the field equations reduce to
\begin{align}
0\,&=\,4\,U'\,\zeta_1'\,+\,\zeta_1''\,+\,\frac{2\,\zeta_1'\,(\overline{\zeta}_1\,\zeta_1'\,+\,\overline{\zeta}_2\,\zeta_2')}{1\,-\,\overline{\zeta}_1\,\zeta_1\,-\,\overline{\zeta}_2\,\zeta_2}\,+\,\frac{3\,g^2}{4}\,\frac{1-3\,\overline{\zeta}_1\,\zeta_1\,-\,2\,\overline{\zeta}_2\,\zeta_2}{1-\overline{\zeta}_1\,\zeta_1\,-\,\overline{\zeta}_2\,\zeta_2}\,, \\ 
0\,&=\,4\,U'\,\zeta_2'\,+\,\zeta_2''\,+\,\frac{2\,\zeta_2'\,(\overline{\zeta}_1\,\zeta_1'\,+\,\overline{\zeta}_2\,\zeta_2')}{1\,-\,\overline{\zeta}_1\,\zeta_1\,-\,\overline{\zeta}_2\,\zeta_2}\,, \\
0\,&=\,3\,(U''\,+\,2\,U'\,U')\,+\,\frac{3}{l^2}\,e^{-2\,U}\, \notag \\
&+2\,\left(\frac{1}{2}\frac{\overline{\zeta'}_1\zeta_1'+\overline{\zeta'}_2\zeta_2'}{1-\overline{\zeta}_1\zeta_1-\overline{\zeta}_2\zeta_2}+\frac{1}{2}\frac{(\overline{\zeta}_1\zeta_1'+\overline{\zeta}_2\zeta_2')(\overline{\zeta'}_1\zeta_1+\overline{\zeta'}_2\zeta_2)}{(1-\overline{\zeta}_1\zeta_1-\,\overline{\zeta}_2\zeta_2)^2}-\frac{3\,g^2}{8}\frac{(1-\overline{\zeta}_2\zeta_2)(2-3\overline{\zeta}_1\zeta_1-2\overline{\zeta}_2\zeta_2)}{(1-\overline{\zeta}_1\zeta_1-\overline{\zeta}_2\zeta_2)^2}\right)\,, \\
0\,&=\,3\,U'\,U'\,+\,\frac{3}{l^2}\,e^{-2\,U}\, \notag \\
&-\,\left(\frac{1}{2}\frac{\overline{\zeta'}_1\zeta_1'+\overline{\zeta'}_2\zeta_2'}{1-\overline{\zeta}_1\zeta_1-\overline{\zeta}_2\zeta_2}+\frac{1}{2}\frac{(\overline{\zeta}_1\zeta_1'+\overline{\zeta}_2\zeta_2')(\overline{\zeta'}_1\zeta_1+\overline{\zeta'}_2\zeta_2)}{(1-\overline{\zeta}_1\zeta_1-\,\overline{\zeta}_2\zeta_2)^2}-\frac{3\,g^2}{8}\frac{(1-\overline{\zeta}_2\zeta_2)(2-3\overline{\zeta}_1\zeta_1-2\overline{\zeta}_2\zeta_2)}{(1-\overline{\zeta}_1\zeta_1-\overline{\zeta}_2\zeta_2)^2}\right)\,.
\end{align}



\end{document}